\documentclass[aps,prl,twocolumn,superscriptaddress]{revtex4-2}
\usepackage[utf8]{inputenc}
\usepackage{color}
\usepackage{amsmath,amssymb,amsthm,mathtools}
\usepackage{xcolor} 
\usepackage{hyperref}
\usepackage{graphicx}
\usepackage{esint}
\usepackage{appendix}
\usepackage{comment}  
\usepackage{braket}
\usepackage{soul}
\usepackage{xcolor}
\setstcolor{red}
\makeatletter
\@ifundefined{textcolor}{}
{%
\definecolor{red}{rgb}{0.75, 0.1, 0.1}
 \definecolor{BLACK}{gray}{0}
 \definecolor{WHITE}{gray}{1}
 \definecolor{RED}{rgb}{1,0,0}
 \definecolor{GREEN}{rgb}{0,1,0}
 \definecolor{BLUE}{rgb}{0,0,1}
 \definecolor{CYAN}{cmyk}{1,0,0,0}
 \definecolor{MAGENTA}{cmyk}{0,1,0,0}
 \definecolor{YELLOW}{cmyk}{0,0,1,0}
}

\newtheorem{theorem}{Theorem}

\newcommand{\Tr}{\operatorname{Tr}}

\newcommand{\Prob}{\mathbb{P}}
\newcommand{\E}{\mathbb{E}}
\newcommand{\mcH}{\mathcal{H}}
\newcommand{\mcS}{\mathcal{S}}
\newcommand{\mcL}{\mathcal{L}}
\newcommand{\eps}{\varepsilon}
\newcommand{\tmix}{t_{\mathrm{mix}}}
\newcommand{\tref}{t_{\mathrm{ref}}}
\newcommand{\tstar}{t_*}
\newcommand{\tworst}{t_{\mathrm{worst}}}
\newcommand{\Prb}{\operatorname{Pr}}

\makeatother

\begin{document}

\title{Typical Mixing and Rare-State Bottlenecks in Open Quantum Systems}

\author{Caisheng Cheng}
\affiliation{Hefei National Laboratory for Physical Sciences at Microscale and Department of Modern
	Physics, University of Science and Technology of China, Hefei, Anhui, China}
\affiliation{Shanghai Branch, CAS Centre for Excellence and Synergetic Innovation Centre in Quantum Information and Quantum Physics, University of Science and Technology of China, Shanghai 201315, China}
\affiliation{Shanghai Research Center for Quantum Sciences, Shanghai 201315, China}

\author{Ruicheng Bao}
\email{ruicheng@g.ecc.u-tokyo.ac.jp}
\affiliation{Department of Physics, Graduate School of Science, The University of Tokyo, Hongo, Bunkyo-ku, Tokyo 113-0033, Japan}

\begin{abstract}
Mixing in open quantum systems is often summarized by a single worst-case time, even though that benchmark can be set by exponentially rare initial states. We show that for broad unstructured ensembles the nonlinear trace-distance relaxation curve itself concentrates around a deterministic mean. For Haar-random pure states this yields fixed-time concentration of the instantaneous trace distance to the steady state, which we term vertical concentration since typical relaxation curves bundle along the distance axis. {Whenever the mean curve crosses the distance threshold with a finite slope, it} converts this vertical concentration into a horizontal concentration of the mixing time, extending typicality from standard physical observables to a fundamentally non-observable dynamical quantity. {This sharp concentration naturally raises the question of how the typical mixing timescale compares to the worst-case benchmark. We show that} in a one-mode tail regime, this separation is controlled by the logarithmic ratio of extremal to typical initial-state overlaps for the slow left eigenoperator.  This rare-state bottleneck law yields a hierarchy that is logarithmic in skin-effect settings, linear for boundary-supported many-body slow modes, and exponential in a protected-sector family where generic states mix rapidly while rare states stagnate. The framework also extends beyond Haar to exact and approximate unitary 2-designs and Hilbert-Schmidt/induced ensembles.
\end{abstract}
\maketitle

\textit{Introduction}---One of the most basic questions in nonequilibrium physics and open quantum systems is how quickly a system loses
memory of how it was prepared. The standard name for this timescale is the mixing time (relaxation time), which is
a key figure of merit in diverse areas, including quantum information and computing \cite{nielsen01quantum,24quantum_glauber,25arxiv_rapid_information}, quantum simulations \cite{11pra_simulations,13njp_simulations,14rmp_simulations,25nature_chen}, state preparation \cite{chen23_preparation,24prr_lin_prep,mi24stable_science,lin25diss_preparation}, and quantum heat engines \cite{06prl_heat_engine,07pre_heat_engine,09pre_heat_engine,19pre_heat_engine,21prl_heat_engines}.

This problem has been extensively studied using rigorous mixing-time bounds  \cite{diaconis1996cutoff,Kastoryano12_cutoff,cubitt15rapidmixing,PhysRevLett.130.060401,24arxiv_optimal_mixing,lin_25cmp_rapidmixing,25cmp_rapid_commute,25arxiv_KMS_rapid,25PRXQuantum_fast_mixing,PhysRevLett.134.140405} and spectral analysis methods \cite{lu2017nonequilibrium,klich2019mpemba,21prl_QMpemba,Goold_24prl,bao_25prl,Ueda_skin_effect,23prb_skin_effect,Mori20resolving,21PRR_mori,21prl_relax_rate_bound,23PRL_mori,24prl_mori,bao25_pre}. However, the puzzle is that mixing is
intrinsically initial-state dependent, yet almost all these studies have focused on finding a timescale that is independent of the initial state to characterize a given dynamical process. The standard choice is the worst-case mixing time, which asks how long one must wait to guarantee convergence no matter how the initial state was chosen. That benchmark is natural if one wants a uniform guarantee. The problem is that the last state to mix can be an exponentially rare and highly structured outlier. A single benchmark then describes a rare bottleneck rather than the relaxation seen by generic preparations.

This is the paradox addressed here. Relaxation in open quantum systems is often strongly uneven. Non-normal Liouvillians can amplify slow left-mode overlaps and generate long tails \cite{Mori20resolving,Ueda_skin_effect,23PRL_mori,24prb_mori}. Mpemba-type suppressions show the opposite possibility. Specially prepared states can evade the same slow sector and relax much faster \cite{ares2025quantum,21prl_QMpemba,Goold_24prl,20prl_raz_precooling,20nature_mpemba,PhysRevLett.131.017101,22prl_mpemba_active,22JPCL_SL,23PRR_bao,ohga_mpemba,PRXLife.2.043019,PhysRevLett.134.107101,hhgj-89gj}. Protected sectors and metastable structures sharpen the contrast further, because most states can relax rapidly while a small sector leaks only on a much longer timescale \cite{PhysRevLett.116.240404,PhysRevResearch.3.033047,25arxiv_rapid_information}. These observations lead to the main question of this work. If worst-case and
generic relaxation differ, can one still define a sharp operational timescale that captures what most initial states do?

{Our recent work highlighted the concentration of linear slow-mode overlaps for generic initial states \cite{bao2026initial}. While this provides a starting point, it does not settle the operational mixing question for two reasons. First,} mixing is governed by the nonlinear trace-distance relaxation curve and by its threshold hitting time. Concentration of linear coefficients does not automatically imply concentration of the curve, and concentration of the curve does not automatically imply concentration of the hitting time. {Second, and importantly, the previous work did not prove the separation between typical and worst-case relaxation timescales, leaving this separation unexplored in concrete physical models and its physical magnitude entirely unquantified.}

Here, we show that such a typical operational scale does exist. Crucially, while previous studies on typicality have predominantly focused on the concentration of physical observables \cite{bao2026initial,06nphy_concentration,06cmp_concentration,06PRL_concentration,09PRE_concentration,09PRL_concentration,23concentration_scipost}, our conceptual advance is to show how typicality can also governs the mixing time, despite it being an inherently nonlinear and non-observable dynamical quantity. We first prove fixed-time vertical concentration: for broad unstructured ensembles, the full trace-distance relaxation curve concentrates around a deterministic mean. When this mean curve crosses the threshold transversely, this vertical concentration is converted into horizontal concentration of the threshold hitting time, thereby giving a high-probability mixing time. We then explain why the usual worst-case scale can nevertheless lie parametrically farther out. {By quantifying how rare states access the slow sector much more strongly than typical ones, we establish a rare-state bottleneck law that produces explicit logarithmic, linear, and exponential typical-versus-worst separations.}

\begin{figure*}[t]
	\centering
	\includegraphics[width=0.98\textwidth]{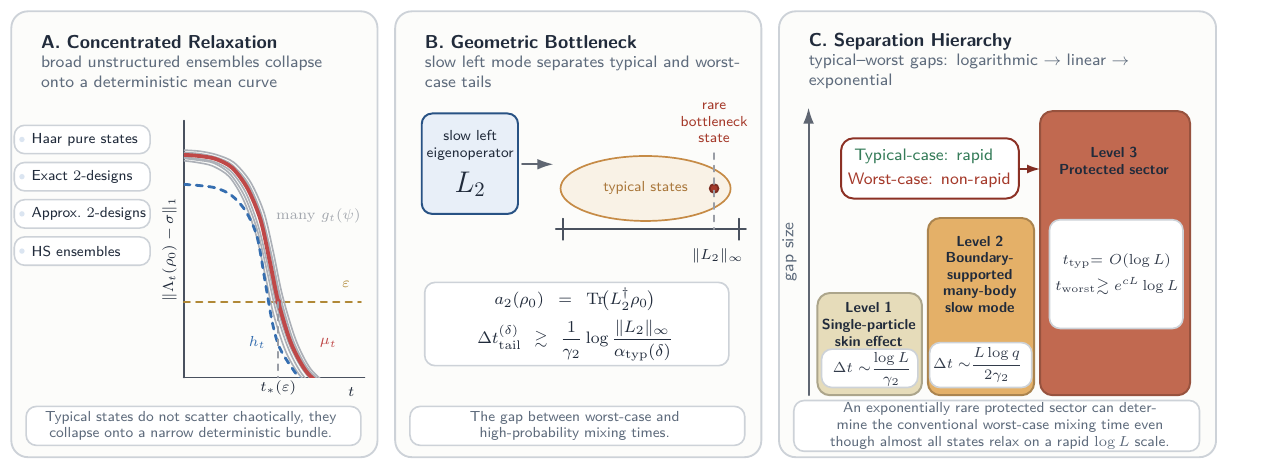}
	\caption{\textbf{Typical-versus-worst mixing.} Panel A shows the concentration theorem. For broad unstructured initial ensembles, the nonlinear relaxation curves \(g_t(\psi)\) form a narrow bundle around the mean \(\mu_t\), while the barycenter trajectory gives a deterministic lower reference. The threshold crossing of \(\mu_t\) defines the bulk time \(t_*=\tstar(\eps)\). Panel B shows the bottleneck estimate. In a one-mode tail, rare states can reach the extremal slow-mode overlap \(\|L_2\|_\infty\), while generic states only see the upper typical scale \(\alpha_{\mathrm{typ}}(\delta)\). This gives \(\Delta t_{\mathrm{tail}}^{(\delta)} \gtrsim \gamma_2^{-1}\log(\|L_2\|_\infty/\alpha_{\mathrm{typ}}(\delta))\). Panel C shows the three resulting branches. The Supplemental Material gives the proofs, the beyond-Haar transfer estimates, and the model details behind the hierarchy.}
	\label{fig:overview}
\end{figure*}

Figure~\ref{fig:overview} summarizes the logic of this work.

\textit{Setup}---
We consider a finite-dimensional Hilbert space $\mcH$ with $d=\dim \mcH$ and a completely positive trace-preserving quantum Markov semigroup $\{\Lambda_t\}_{t\ge 0}$ generated by a Lindbladian $\mcL$,
\begin{equation}
	\frac{d}{dt}\rho(t)=\mcL(\rho(t)),
	\qquad
	\Lambda_t=e^{t\mcL}.
	\label{eq:qms}
\end{equation}
Throughout, we assume that $\mcL$ admits a unique stationary state $\sigma\in\mcS(\mcH)$, so $\mcL(\sigma)=0$ and $\Lambda_t(\sigma)=\sigma$ for all $t\ge 0$.

For an initial state $\rho_0$ and accuracy threshold $\eps\in(0,1)$, the trace-distance mixing time is
\begin{equation}
	\tmix(\rho_0,\eps):=
	\inf\left\{
	t\ge 0:\ \|\Lambda_t(\rho_0)-\sigma\|_1\le \eps
	\right\}.
	\label{eq:tmix}
\end{equation}
We begin with Haar-random pure states. This case is simple enough to prove concentration for the full nonlinear curve, and it is the reference ensemble for the model estimates below. The later arguments do not rely on Haar randomness itself, but on fixed-time curve control and a high-probability slow-overlap scale. In Sec.~S6 of
the Supplemental Material (SM)  \cite{supplemental_material}, these two inputs are verified for exact and approximate unitary \(2\)-designs and for Hilbert--Schmidt/induced ensembles \cite{supplemental_material}. For $\ket{\psi}\in\mcH$, define
\begin{equation}
	g_t(\psi):=
	\left\|\Lambda_t\!\bigl(\ket{\psi}\!\bra{\psi}\bigr)-\sigma\right\|_1,
	\qquad
	\mu_t:=\E_{\mathrm{Haar}}[g_t(\psi)].
	\label{eq:gandmu}
\end{equation}
The deterministic crossing time of the mean curve is
\begin{equation}
	\tstar(\eps):=\inf\{t\ge 0:\ \mu_t\le \eps\}.
	\label{eq:ttyp}
\end{equation}
A typical mixing scale should be attached to the bulk of initial states rather than to an adversarial outlier. We compare the mean crossing \(t_*(\eps)\) with a high-probability mixing time and with the adversarial benchmark. The high-probability scale is
\begin{equation}
	t_{\mathrm{typ}}^{(\delta)}(\eps):=
	\inf\bigl\{
	t\ge 0:\ 
	\Prob_{\mathrm{Haar}}\!\bigl(
	\tmix(\ket{\psi}\!\bra{\psi},\eps)\le t
	\bigr)\ge 1-\delta
	\bigr\},
	\label{eq:quantiletyp}
\end{equation}
and the adversarial benchmark is
\begin{equation}
	\tworst(\eps):=
	\sup_{\|\psi\|=1}\tmix(\ket{\psi}\!\bra{\psi},\eps).
	\label{eq:worstdef}
\end{equation}

\textit{Vertical concentration of relaxation curves}---
The relevant object is the curve \(g_t(\psi)\), not just a mode coefficient or a gap. Different initial states could in principle follow different trace-distance histories. If that happened, the quantile time in \eqref{eq:quantiletyp} would have no simple deterministic reference. The theorem below rules out this obstruction for Haar-random pure states at each fixed time.

\begin{theorem}[Fixed-time concentration of the relaxation curve]
	\label{thm:vertical}
	Let $\{\Lambda_t\}_{t\ge 0}$ be a CPTP quantum Markov semigroup with unique stationary state $\sigma$, and let $g_t$, $\mu_t$, and \(\tstar\) be defined by \eqref{eq:gandmu}--\eqref{eq:ttyp}. Then there exists an absolute constant $c>0$ such that for every $t\ge 0$ and every $\eta>0$,
	\begin{equation}
		\Prob_{\mathrm{Haar}}\!\left(|g_t(\psi)-\mu_t|\ge \eta\right)
		\le
		2e^{-c d\eta^2}.
		\label{eq:verticaltail}
	\end{equation}
	\par
	Assume in addition that $\mu_0>\eps>\lim_{t\to\infty}\mu_t$, and set \(t_*:=\tstar(\eps)\). Then $\mu_{t_*}=\eps$, and for every $\eta>0$,
	\begin{align}
		\Prob_{\mathrm{Haar}}\!\left(
		\tmix(\ket{\psi}\!\bra{\psi},\eps+\eta)\le t_*
		\right)
		&\ge 1-2e^{-c d\eta^2},
		\label{eq:onesidedright}
		\\
		\Prob_{\mathrm{Haar}}\!\left(
		\tmix(\ket{\psi}\!\bra{\psi},\eps-\eta)\ge t_*
		\right)
		&\ge 1-2e^{-c d\eta^2}.
		\label{eq:onesidedleft}
	\end{align}
\end{theorem}

Theorem~\ref{thm:vertical} gives the deterministic reference we need. At fixed time, the random nonlinear curve lies near \(\mu_t\). Generic initial states therefore do not produce unrelated relaxation profiles.

\textit{Horizontal concentration of mixing times}---Fixed-time concentration is still not a statement about the same-threshold mixing time. A small vertical error near a flat crossing can shift the hitting time by a large amount. The next theorem adds the local slope input needed to turn the fixed-time bundle into a crossing-time bundle.
\begin{theorem}[Concentration of the mixing time from a transverse crossing]
	\label{thm:horizontal}
	Fix $\eps\in(0,1)$ and assume $\mu_0>\eps>\lim_{t\to\infty}\mu_t$. Let \(t_*:=\tstar(\eps)\). Suppose there exist constants $m>0$ and $\delta_0>0$ such that
	\begin{equation}
		\mu_{t_*-u}\ge \eps+m u,
		\qquad
		\mu_{t_*+u}\le \eps-m u,
		\qquad
		0\le u\le \delta_0.
		\label{eq:transverse}
	\end{equation}
	Then for every $\eta\in(0,m\delta_0]$,
	\begin{equation}
		\Prob_{\mathrm{Haar}}\!\left(
		\left|\tmix(\ket{\psi}\!\bra{\psi},\eps)-t_*\right|>\frac{\eta}{m}
		\right)
		\le
		4e^{-c d\eta^2},
		\label{eq:horiztail}
	\end{equation}
	where $c$ is the constant from Theorem~\ref{thm:vertical}.
\end{theorem}

Theorem~\ref{thm:horizontal} is the operational core of this work. As mentioned before, Theorem~\ref{thm:vertical} gives vertical concentration; however, concentration of the same-threshold mixing time needs a local inverse step at the crossing. The slope condition~\ref{eq:transverse} provides exactly that step. It is a finite-regime expression of a transverse crossing. Near \(t_*(\varepsilon)\), the mean curve is not locally flat at the threshold. Equivalently, a vertical error of size \(\eta\) can move the crossing time by at most \(\eta/m\). If the curve were flat, the same vertical fluctuation could produce a much larger horizontal uncertainty. In this sense, Theorem~\ref{thm:horizontal} is Theorem~\ref{thm:vertical} plus a transverse-crossing input. The constant $m$ measures the local steepness of the mean curve at the chosen threshold. In a one-mode tail, where $\mu_t \approx A e^{-\gamma t}$, this gives $m \approx \gamma \eps$ so the corresponding semilog slope is $\gamma$. For an isolated real slow mode, this is the spectral gap. Outside that regime, $m$ should be read only as a local effective slope. The proof
is given in Sec.~S3 of SM \cite{supplemental_material}. The fixed-rate discussions are given in Sec.~S8.

Equivalently, under the transverse-crossing hypothesis, the operational quantile scale \(t_{\mathrm{typ}}^{(\delta)}(\eps)\) lies within \(O\!\left(\frac{1}{m}\sqrt{\log(1/\delta)/d}\right)\) of the deterministic crossing time \(t_*(\eps)\). It is worth noting that Theorem~\ref{thm:horizontal} is only a local sufficient criterion for identifying the quantile scale \(t_{\mathrm{typ}}^{(\delta)}(\varepsilon)\) with the deterministic crossing time \(t_*(\varepsilon)\), so the quantile definition of \(t_{\mathrm{typ}}^{(\delta)}(\varepsilon)\) stands independently of the theorem. Theorem~\ref{thm:horizontal} is a fixed-threshold statement. It identifies a typical mixing scale at a given accuracy \(\eps\) from a transverse crossing of the mean curve at that same threshold. A uniform claim over a whole threshold range would require an additional slope input that remains valid across that range.

\begin{figure*}[t]
	\centering
	\includegraphics[width=0.66\textwidth]{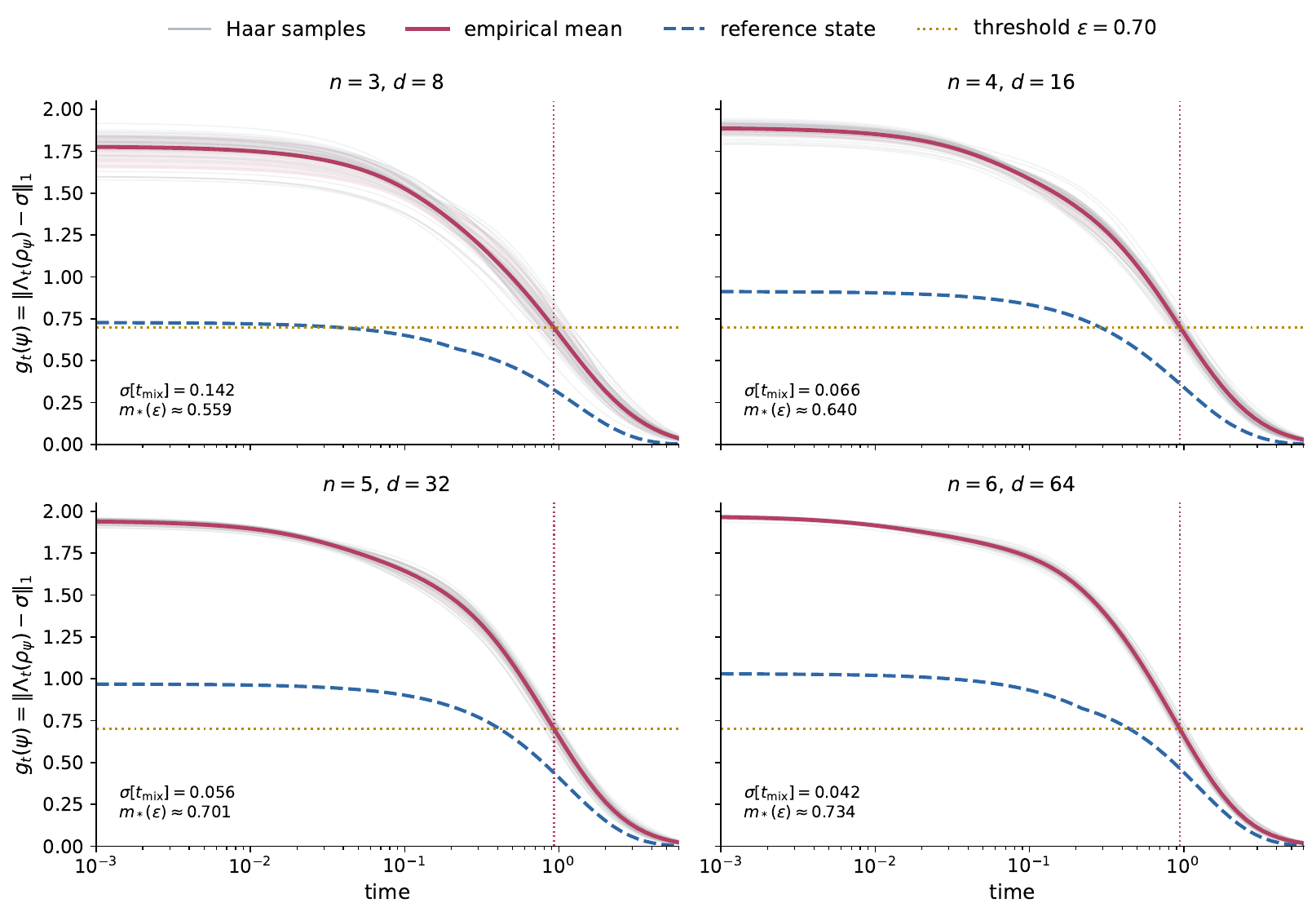}\hfill
	\raisebox{0.0058\textheight}{
	\includegraphics[width=0.33\textwidth]{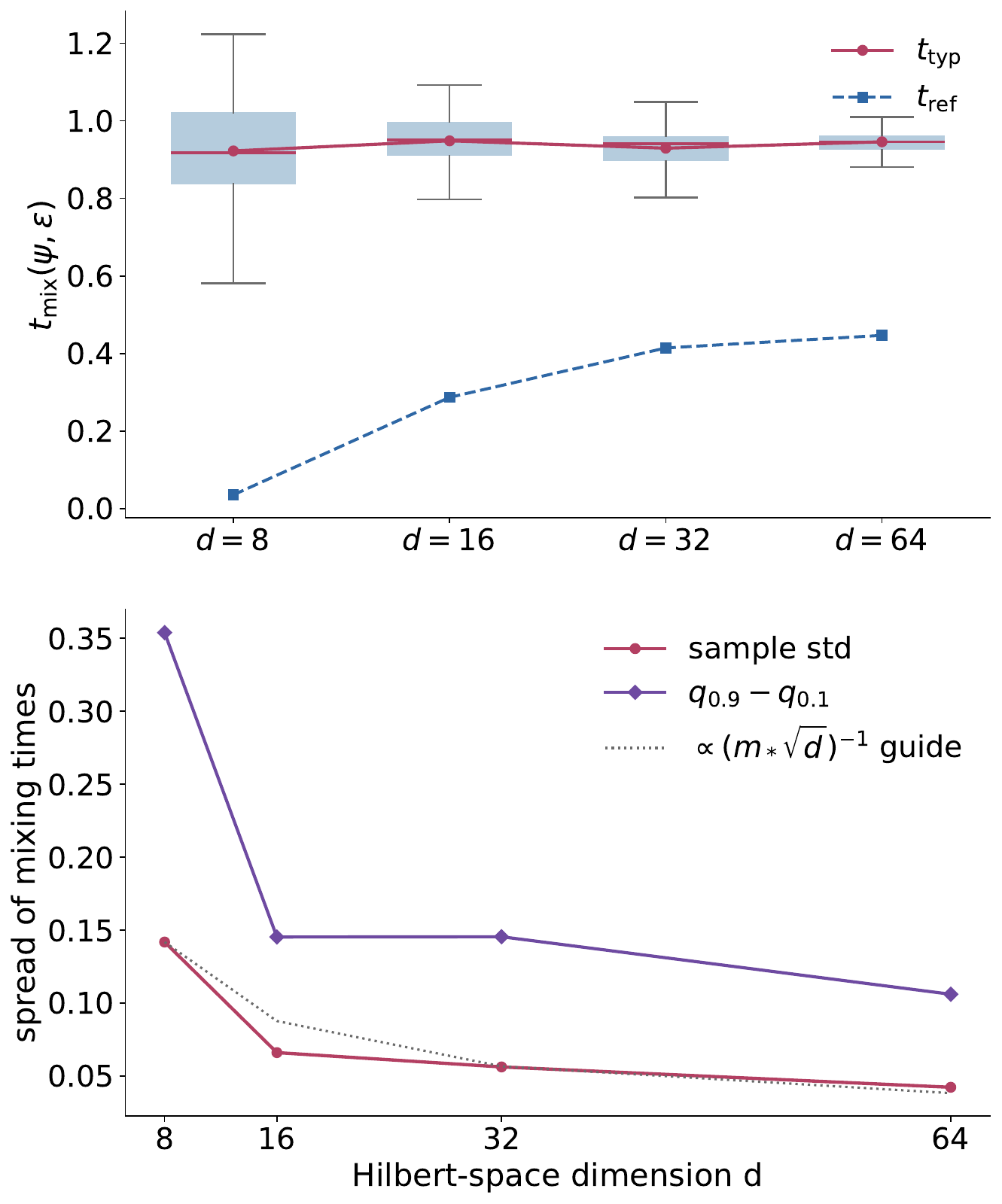}}
	\caption{\textbf{Davies test for vertical and horizontal concentration.} The left panel shows bundle panels for a local many-body Davies semigroup with product Gibbs fixed point, using the supplementary Davies model of Sec.~S4. Gray curves are Haar-random trajectories, red is the empirical mean, blue dashed is the barycenter benchmark $h_t=\left\|\Lambda_t\left(\rho_{\text {ref }}\right)-\sigma\right\|_1$, generated by the ensemble barycenter $\rho_{\text {ref }}=\mathbb{E}_{\text {Haar }}[|\psi\rangle\langle\psi|]$. It is included as a deterministic lower reference for the bulk bundle. The threshold crossing becomes visibly sharper as the Hilbert-space dimension grows. The right panel shows fixed-threshold mixing-time statistics for the same model. The empirical spread of \(\tmix(\psi,\eps)\) decreases with dimension, which is the finite-size signature of Theorem~\ref{thm:horizontal}.}
	\label{fig:davies-main}
\end{figure*}

Figure~\ref{fig:davies-main} visualizes the vertical-to-horizontal mechanism. In the Davies benchmark, Haar-random relaxation curves already form a narrow bundle at fixed time, while the deterministic reference trajectory remains below the bulk bundle (see the End Matter). More importantly, the threshold crossings themselves also cluster as the Hilbert-space dimension grows. The right panel turns this into the operational statement. Both the standard deviation and the interquantile width of \(\tmix(\psi,\eps)\) decrease with dimension, while the crossing scale itself remains \(O(1)\). The concentration is therefore not a trivial time rescaling. It is genuine narrowing of the mixing-time distribution.

\textit{Rare-State Bottleneck Law}---
The concentration theorems solve only the first half of the paradox. They identify a typical scale. The remaining question is why the conventional worst-case benchmark can still be much larger. The answer is geometric. In the one-mode tail regime, rare states can attain slow-mode overlaps that are unavailable to the typical bulk.

Here we assume \(\mcL\) is diagonalizable. Let \(\{R_k\}\) and \(\{L_k\}\) be biorthogonal right and left eigenoperators satisfying
$
	\mcL(R_k)=\lambda_k R_k,
	\;
	\mcL^{\dagger}(L_k)=\overline{\lambda_k}L_k,
	\;
	\Tr\!\left(L_k^{\dagger}R_h\right)=\delta_{kh}.
$
Then
$
	\Lambda_t(\rho_0)-\sigma
	=
	\sum_{k\ge 2} e^{\lambda_k t} a_k(\rho_0) R_k,
	\;\text{where}\;
	a_k(\rho_0):=\Tr(L_k^\dagger \rho_0).
$
We focus on the clean one-mode tail regime, where the \(\eps\)-regime of interest lies late enough that the real slowest mode dominates \cite{PhysRevResearch.3.033047,PhysRevLett.116.240404}. This is the most convenient regime for analysis because the tail is then controlled by a single overlap scale. This should not be interpreted as the only regime in which a worst-versus-typical separation
can appear. 
The fixed-rate skin numerics in the SM \cite{supplemental_material} Sec.~S8 show that a worst-versus-typical separation already appears at larger thresholds before the fully clean one-mode tail. In that regime, the crossing samples an effective slow sector with localized support rather than a single eigenmode, and the logarithmic trend survives with visible subleading-mode corrections because rare and typical initial states still have parametrically different overlaps with that sector. Let \(\alpha_{\mathrm{typ}}(\delta)\) denote a \((1-\delta)\)-high-probability upper scale for the slow overlap \(|a_2(\psi)|\) under Haar sampling. So we have
\begin{equation}
	\Delta t_{\mathrm{tail}}^{(\delta)}(\eps):=
	\tworst(\eps)-t_{\mathrm{typ}}^{(\delta)}(\eps)
	\gtrsim
	\frac{1}{\gamma_2}\log\!\left(
	\frac{\|L_2\|_\infty}{\alpha_{\mathrm{typ}}(\delta)}
	\right).
	\label{eq:gapheuristic}
\end{equation}
Equation~\eqref{eq:gapheuristic} is the bottleneck law. Its origin is simple once the late-time dynamics is resolved spectrally. In the one-mode regime, the relaxation tail is governed by the slow overlap \(|a_2(\psi)|\) multiplied by the common decay factor \(e^{-\gamma_2 t}\). The slowest admissible initial state can align with the left mode as strongly as \(\|L_2\|_\infty\), whereas Haar-typical initial states see only the high-probability scale \(\alpha_{\mathrm{typ}}(\delta)\). Solving the same threshold condition for these two overlap scales immediately yields a logarithmic time gap.

The key point is that this law is quantile-based. To control \(t_{\mathrm{typ}}^{(\delta)}(\eps)\), one only needs the event of anomalously large slow-mode overlap to have probability at most \(\delta\). Those rare large-overlap states are exactly what the worst-case benchmark keeps and \(t_{\mathrm{typ}}^{(\delta)}(\eps)\) discards, while states with anomalously small overlap, {including strong Mpemba-like states with \(a_2\simeq0\),} only mix faster \cite{klich2019mpemba,21prl_QMpemba,Goold_24prl,ares2025quantum}. {If a high-probability ensemble has its late tail controlled by a subleading mode because the slowest-mode overlap is typically suppressed, the same quantile-overlap argument can be repeated with the next relevant slow channel \(a_k\) and its decay rate \(\gamma_k=-\mathrm{Re}\,\lambda_k\).} In the first two applications below, this law is used in fixed-rate baselines. The growth of the gap is then attributed to the extremal-to-typical overlap ratio of the same slow sector. If \(\gamma_2(L)\) is also allowed to close, the same numerator is multiplied by the common factor \(\gamma_2(L)^{-1}\). That enlarges the absolute gap, but it does not sharpen the overlap mechanism itself. The rigorous one-mode theorem is given in the Supplemental Material \cite{supplemental_material} Sec.~S5. The fixed-rate checks are discussed in Sec.~S8.

\textit{A Hierarchy of Typical-Worst Separations}---
Equation~\eqref{eq:gapheuristic} reduces the hierarchy question to one geometric ratio. So how large can the slow-mode overlap be, and how large is it for almost all initial states? The models below give three answers. A fixed-rate skin model gives a logarithmic gap, a fixed-rate boundary-supported many-body model gives a linear gap, and a protected-sector model gives an exponential separation. The physical pictures are summarized in the End Matter.

For the skin branch, the ambient one-particle dimension is \(L\). In the fixed-rate single-particle skin-effect model of Supplemental Material \cite{supplemental_material} Sec.~S7.1, the slow left mode is boundary localized. The extremal overlap stays \(O(1)\), while the Haar-typical boundary weight is only algebraically small. The typical--worst gap is therefore logarithmic in \(L\).
\begin{equation}
	\tworst(\eps)-t_{\mathrm{typ}}^{(\delta)}(\eps)
	\ge
	\frac{\log L}{\gamma_2}+O(1).
	\label{eq:main-skinlog}
\end{equation}
For the boundary-supported many-body branch, the same overlap geometry is placed in a Hilbert space of dimension \(q^L\). In \cite{supplemental_material} Sec.~S7.2, the slow observable lives on a fixed boundary block while the bulk dimension grows exponentially. The typical overlap is suppressed as \(q^{-L/2}\), giving
\begin{equation}
	\tworst(\eps)-t_{\mathrm{typ}}^{(\delta)}(\eps)
	\ge
	\frac{L\log q}{2\gamma_2}+O(1).
	\label{eq:main-boundarylinear}
\end{equation}
The logarithmic and linear branches are fixed-rate baselines. They are meant to isolate the overlap geometry of the same slow sector. SM \cite{supplemental_material} Sec.~S8 discusses what changes when the slow rate is allowed to close with \(L\).

The protected-sector branch uses a different mechanism. In the dephasing-plus-jump family of Supplemental Material \cite{supplemental_material} Sec.~S7.3, one protected basis state leaks only at rate \(\eta_L=e^{-cL}\), while the remaining \(d-1=q^L-1\) states mix on an \(O(1)\) scale. A Haar-random state barely occupies the protected state. High-probability mixing is therefore fast, while a specially prepared initial state follows the slow leakage time. In particular,
\begin{equation}
t_{\mathrm{typ}}^{(\delta)}(\eps_L)=O(\log L),
\qquad
\tworst(\eps_L)\ge \Omega\bigl(e^{cL}\log L\bigr)
\end{equation}
for every fixed \(\delta\in(0,1)\) and every inverse-polynomial threshold sequence \(\eps_L=L^{-m}\). Logical-sector and local-qubit versions are given in SM \cite{supplemental_material} Secs.~S7.4 and S7.5 \cite{25arxiv_rapid_information,RevModPhys.88.045005,dennis2002topological,RevModPhys.87.307}.

\textit{Discussion and Outlook}---
Open-system mixing need not be summarized by a single adversarial timescale. For broad unstructured ensembles, the full nonlinear trace-distance relaxation curve concentrates around a deterministic mean, and a transverse crossing turns that concentration into a typical operational mixing scale. At the same time, the conventional worst-case scale can lie much farther out, because rare initial states can couple to the slow sector much more strongly than the typical bulk.

This distinction matters whenever mixing time is used as a proxy for algorithmic cost, as in dissipative Gibbs sampling, thermalization estimates, and ground-state preparation \cite{temme2010chi,chen23_preparation,chen23efficient,25nature_chen,lin25diss_preparation,zhan2026rapid}. Worst-case mixing remains the correct notion when adversarial guarantees are required. For randomized or weakly structured initializations, however, it can misidentify the operational timescale, and in protected-sector settings it can do so exponentially. From this viewpoint, skin-effect enhancement, Mpemba-type suppression, protected sectors, and logical-sector relaxation are not disconnected anomalies. They are different ways in which slow-sector geometry redistributes weight between generic states and rare outliers \cite{Mori20resolving,Ueda_skin_effect,23PRL_mori,24prb_mori,21prl_QMpemba,22pra_qmpemba_carollo,Goold_24prl,bao_25prl,PhysRevLett.131.017101,teza2026speedups,ares2025quantum,PhysRevLett.116.240404,25arxiv_rapid_information}. More generally, the relevant relaxation law requires dynamical input beyond the bare Liouvillian gap. This is already visible in skin-type and gap-discrepancy settings, and even in detailed-balance families much stronger thermalization results emerge once additional structure is imposed \cite{25cmp_rapid_commute,kastoryano2013quantum}.

{It is worth noting that the Haar ensemble is mainly an analytic reference, not a requirement to implement full Haar randomization. Sec.~S6 of the Supplemental Material shows that the same slow-overlap and quantile logic extends to exact \(2\)-designs, approximate \(2\)-designs with explicit \(O(\varepsilon_2)\) errors, and Hilbert--Schmidt/induced ensembles, where \(d_B\) further suppresses the typical slow-overlap scale \cite{supplemental_material}. The full-curve tail bound is then polynomial rather than Haar-L\'evy exponential. Experimentally, one may replace Haar unitaries by Clifford \(2\)-designs or approximate random-circuit designs. Recent constructions give \(O(\log n)\) depth without ancillas, or \(O(\log\log(n/\varepsilon_2))\) depth for fixed \(k=2\) using \(\widetilde O(n)\) ancillas \cite{dankert2009exact,schuster2025random,cui2025unitary}.} Furthermore, our results do not establish a quantum--classical complexity separation. They suggest instead an initialization-aware view of dissipative complexity, in line with questions about typical quantum advantage \cite{huang2025vast,lin25diss_preparation}. A natural extension is to treat multimode and non-diagonalizable slow sectors with the same quantile-based logic, and to develop diagnostics for distinguishing statistically dominant slow channels from genuinely rare bottlenecks in microscopic models.

\textit{Acknowledgments---}We are grateful to Chaoyang Lu, Zhongxia Shang, Suyang Lin and Yijia Cheng for useful discussions. C. C. is supported by the National Natural Science Foundation of China (No.124B100020).  
R.~B. is supported by JSPS KAKENHI Grant No.\ 25KJ0766.

\textit{Data availability statement---}The code used to generate all the figures is openly available at \footnote{https://github.com/Ustcmilkquamtum/code-for-Typical-Mixing-and-Rare-State-Bottlenecks-in-Open-Quantum-Systems.git}. 

\bibliography{refs}

\begin{center}
{\large\textbf{End Matter}}
\end{center}

\begin{figure*}[t]
	\centering
	\includegraphics[width=0.92\textwidth]{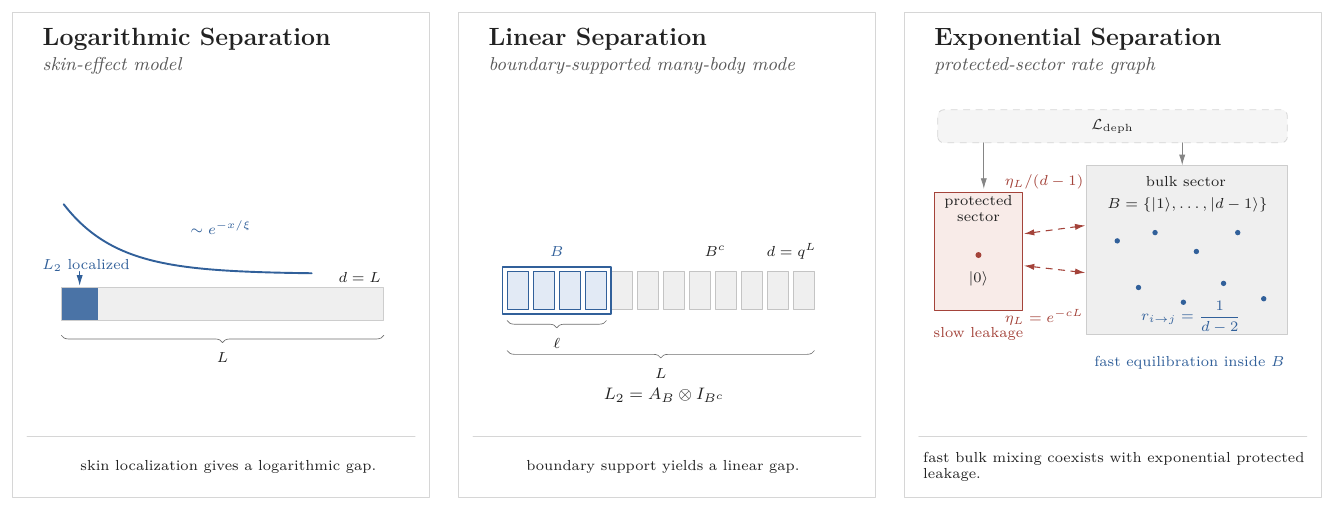}
	\caption{Left: fixed-rate single-particle skin baseline. Boundary localization keeps the extremal slow overlap of order one, while the typical overlap decays only algebraically in the effective one-particle dimension, which yields a logarithmic gap. Middle: fixed-rate boundary-supported many-body baseline. The slow operator remains supported on an $O(1)$-site boundary algebra, but the ambient Hilbert space grows as $q^L$, so the typical overlap is diluted by $q^{-L / 2}$, which yields a linear gap. Right: protected-sector branch. The high-probability sector relaxes through the fast bulk, while the worst-case benchmark is set by exponentially slow leakage from a rare protected configuration.}
	\label{fig:hierarchy}
\end{figure*}
For the full Haar ensemble, the barycenter is the maximally mixed state. This barycenter defines a deterministic reference trajectory,
$\rho_{\mathrm{ref}}=\mathbb{E}_{\mathrm{Haar}}[|\psi\rangle\langle\psi|], \quad h_t:=\left\|\Lambda_t\left(\rho_{\mathrm{ref}}\right)-\sigma\right\|_1, \quad t_{\mathrm{ref}}(\varepsilon):=\inf \left\{t \geq 0: h_t \leq \varepsilon\right\}$. Because the trace norm is convex,
\begin{align}
	h_t&=\left\|\mathbb{E}_{\text {Haar }}\left[\Lambda_t(|\psi\rangle\langle\psi|)-\sigma\right]\right\|_1 \notag\\
	&\leq \mathbb{E}_{\text {Haar }}\left[\left\|\Lambda_t(|\psi\rangle\langle\psi|)-\sigma\right\|_1\right]=\mu_t,
\end{align}
Hence $\tref(\eps)\le \tstar(\eps)$. So the barycenter trajectory is a deterministic lower benchmark for the concentrated bulk.

In dissipative state-preparation algorithms such as Zhan \emph{et al.} \cite{zhan2026rapid}, the maximally mixed state is used as a convenient concrete initialization, and convergence is often monitored through energy-based or fidelity-based surrogate mixing times for that chosen start state. The same maximally mixed state is indeed a suitable choice. From our perspective, it is the barycenter of the full Haar ensemble. The statement $h_t\le \mu_t$ therefore does not say that the maximally mixed evolution is typical. It says only that the barycenter supplies a simple deterministic reference curve that lies below the concentrated bulk curve.

This viewpoint naturally suggests a broader initialization-aware notion of mixing beyond trace distance. Let $D(\rho,\sigma)$ be any application-level discrepancy functional relative to the target state $\sigma$, for example a trace-distance error, an energy error, a fidelity deficit, a local-observable error, or a logical-sector error. For an initial state $\rho_0$, define
\[
\tmix^{D}(\rho_0,\eps):=\inf\{t\ge 0:\ D(\Lambda_t(\rho_0),\sigma)\le \eps\}.
\]
For an initialization ensemble $\nu$, the corresponding high-probability benchmark is
\[
t_{\mathrm{typ},\nu}^{D,(\delta)}(\eps)
:=
\inf\Bigl\{t\ge 0:\ \Prb_{\rho_0\sim \nu}\bigl[\tmix^{D}(\rho_0,\eps)\le t\bigr]\ge 1-\delta\Bigr\},
\]
while the adversarial benchmark is
\[
\tworst^{D}(\eps):=\sup_{\rho_0}\,\tmix^{D}(\rho_0,\eps).
\]
When $D$ is the trace distance, this recovers the operational notion studied in the main text. When $D$ is an observable-specific surrogate, these definitions quantify the relaxation scale seen by most initial states in the ensemble relevant to the application.

For dissipative ground-state preparation, one natural choice is a normalized energy error. Another is a fidelity deficit relative to the target ground state, exactly in the spirit of Zhan \emph{et al.} \cite{zhan2026rapid}. The resulting typical energy-mixing or typical fidelity-mixing time would not replace the trace-distance benchmark developed here. It would instead complement that benchmark by asking how quickly most accessible initial states converge in the quantity that the algorithm or experiment actually monitors. Developing a systematic theory that relates such observable-specific typical benchmarks to the universal trace-distance typical benchmark is, in our view, a natural next step.

The hierarchy in Fig.~\ref{fig:hierarchy} follows the same distinction. In the skin and boundary branches, typical and worst states probe the same slow sector with different overlap scales. In the protected-sector branch, typical states almost never enter the slow sector. The corresponding constructions are given in Supplemental Material \cite{supplemental_material} Secs.~S7 and S8.

The skin and boundary branches should be read together with Supplemental Material \cite{supplemental_material} Sec.~S8, where the examples are kept at fixed rate to separate overlap geometry from a common dilation by \(\gamma_2(L)^{-1}\). The protected-sector branch is different. Its separation comes from exponentially small Haar weight in the protected sector, so typical mixing is governed by the fast bulk while the worst-case scale follows slow leakage.

\end{document}


\title{Supplemental Material for ``Typical Mixing and Rare-State Bottlenecks in Open Quantum Systems''}

\author{Caisheng Cheng}
\affiliation{Hefei National Laboratory for Physical Sciences at Microscale and Department of Modern
	Physics, University of Science and Technology of China, Hefei, Anhui, China}
\affiliation{Shanghai Branch, CAS Centre for Excellence and Synergetic Innovation Centre in Quantum Information and Quantum Physics, University of Science and Technology of China, Shanghai 201315, China}
\affiliation{Shanghai Research Center for Quantum Sciences, Shanghai 201315, China}

\author{Ruicheng Bao}
\email{ruicheng@g.ecc.u-tokyo.ac.jp}
\affiliation{Department of Physics, Graduate School of Science, The University of Tokyo, Hongo, Bunkyo-ku, Tokyo 113-0033, Japan}

\maketitle

This supplement contains the proofs and model estimates used in the main text. We start with the Haar concentration argument and the crossing-time estimate. We then give the finite-size Davies check behind Fig.~2, the one-mode bottleneck bound, and the ensemble-transfer estimates. The last part records the skin, boundary, protected-sector, and logical-sector examples, including the fixed-rate numerical checks used for the separation hierarchy.

\tableofcontents

\newpage

\vspace{12pt}

\section{Haar Moments and Concentration Lemmas}
\label{app:haar}

\begin{lemma}[Haar moments of quadratic forms {\cite[Corollary~13]{mele2024introduction}}]
	\label{lem:haarmoments}
	Let $O=O^\dagger\in \mathcal{L}(\mathbb{C}^d)$ and let $\ket{\psi}$ be Haar-random. Then
	\begin{align}
		\E_{\mathrm{Haar}}\!\left[\bra{\psi}O\ket{\psi}\right]
		=
		\frac{\Tr(O)}{d},\qquad
		\E_{\mathrm{Haar}}\!\left[\bra{\psi}O\ket{\psi}^2\right]
		=
		\frac{\Tr(O)^2+\Tr(O^2)}{d(d+1)}.
		\label{eq:secondmoment}
	\end{align}
	Hence
	\begin{align}
		\operatorname{Var}_{\mathrm{Haar}}\!\left(\bra{\psi}O\ket{\psi}\right)
		&=
		\frac{\Tr(O^2)-\Tr(O)^2/d}{d(d+1)}.
		\label{eq:varmoment}
	\end{align}
\end{lemma}

\begin{proof}
	The one-fold twirl gives the mean as a trace identity,
	\begin{align*}
		\E_{\mathrm{Haar}}\!\left[\bra{\psi}O\ket{\psi}\right]
		=
		\Tr\!\left(O\,\E_{\mathrm{Haar}}[\ket{\psi}\!\bra{\psi}]\right)
		=
		\Tr\!\left(O\,\frac{\Id}{d}\right)
		=
		\frac{\Tr(O)}{d}.
	\end{align*}
	The two-fold twirl is
	$
	\E_{\mathrm{Haar}}\!\left[(\ket{\psi}\!\bra{\psi})^{\otimes 2}\right]
	=
	\frac{\Id+\mathbb{F}}{d(d+1)},
	$
	where $\mathbb{F}$ is the swap operator. Tracing this identity against \(O\otimes O\) gives the complete second-moment calculation,
	\begin{align*}
		\E_{\mathrm{Haar}}\!\left[\bra{\psi}O\ket{\psi}^2\right]
		=
		\Tr\!\left[
		(O\otimes O)
		\E_{\mathrm{Haar}}\!\left((\ket{\psi}\!\bra{\psi})^{\otimes 2}\right)
		\right]
		=
		\frac{\Tr[(O\otimes O)(\Id+\mathbb{F})]}{d(d+1)}
		=
		\frac{\Tr(O)^2+\Tr(O^2)}{d(d+1)}.
	\end{align*}
	The variance formula follows by substituting \eqref{eq:secondmoment} into the centered second moment,
	\begin{align*}
		\operatorname{Var}_{\mathrm{Haar}}\!\left(\bra{\psi}O\ket{\psi}\right)
		=
		\E_{\mathrm{Haar}}\!\left[\bra{\psi}O\ket{\psi}^2\right]
		-
		\left(
		\E_{\mathrm{Haar}}\!\left[\bra{\psi}O\ket{\psi}\right]
		\right)^2
		=
		\frac{\Tr(O)^2+\Tr(O^2)}{d(d+1)}
		-
		\frac{\Tr(O)^2}{d^2}
		=
		\frac{\Tr(O^2)-\Tr(O)^2/d}{d(d+1)}.
	\end{align*}
\end{proof}

\begin{lemma}[Trace norm of pure-state projector differences]
	\label{lem:ranktwo}
	For unit vectors $u,v\in\mathbb{C}^d$,
	\begin{equation}
		\left\|\ket{u}\!\bra{u}-\ket{v}\!\bra{v}\right\|_1
		=
		2\sqrt{1-|\braket{u}{v}|^2}
		\le
		2\|u-v\|_2.
		\label{eq:ranktwo}
	\end{equation}
\end{lemma}

\begin{proof}
	The Hermitian operator
	\(\ket{u}\!\bra{u}-\ket{v}\!\bra{v}\)
	has rank at most two and trace zero. Its squared Hilbert--Schmidt norm is
	\begin{align*}
		\Tr\!\left[
		\left(\ket{u}\!\bra{u}-\ket{v}\!\bra{v}\right)^2
		\right]
		=
		\Tr(\ket{u}\!\bra{u})
		+
		\Tr(\ket{v}\!\bra{v})
		-
		2\,\Tr(\ket{u}\!\bra{u}\ket{v}\!\bra{v})
		=
		2\bigl(1-|\braket{u}{v}|^2\bigr).
	\end{align*}
	The two nonzero eigenvalues are therefore
	\(
	\pm \sqrt{1-|\braket{u}{v}|^2}
	\),
	which proves the equality in \eqref{eq:ranktwo}. The Euclidean bound is 
	\begin{align*}
		1-|\braket{u}{v}|^2
		=
		(1-|\braket{u}{v}|)(1+|\braket{u}{v}|)
		\le
		2\bigl(1-|\braket{u}{v}|\bigr)
		\le
		2\bigl(1-\Re\braket{u}{v}\bigr)
		=
		\|u-v\|_2^2.
	\end{align*}
	Taking square roots gives the bound.
\end{proof}

\begin{lemma}[L\'evy's lemma {\cite{ledoux2001concentration}}]
	\label{lem:levy}
	Let $f$ be a real-valued $L$-Lipschitz function on the unit sphere of $\mathbb{C}^d$, equipped with the Euclidean metric $\|\cdot\|_2$. Then there exists an absolute constant $c>0$ such that for every $\eta>0$,
	\begin{equation}
		\Prob_{\mathrm{Haar}}\!\left(|f(\psi)-\E_{\mathrm{Haar}}[f]|\ge \eta\right)
		\le
		2e^{-cd\eta^2/L^2}.
		\label{eq:levy}
	\end{equation}
\end{lemma}

\section{Proof of Theorem~1}
\label{app:vertical}

\begin{proof}[Proof of Theorem~1]
	We begin with the Lipschitz estimate for the relaxation curve. For a unit vector \(\psi\), write
	$
	\rho_\psi:=\ket{\psi}\!\bra{\psi}.
	$
	For fixed \(t\ge 0\), regard
	$
	g_t(\psi):=
	\left\|
	\Lambda_t(\rho_\psi)-\sigma
	\right\|_1
	$
	as a function of the initial vector \(\psi\).

	For unit vectors \(u,v\), let \(\rho_u:=\ket{u}\!\bra{u}\) and \(\rho_v:=\ket{v}\!\bra{v}\). The Lipschitz estimate is 
	\begin{align}
		|g_t(u)-g_t(v)|
		&=
		\left|
		\left\|\Lambda_t(\rho_u)-\sigma\right\|_1
		-
		\left\|\Lambda_t(\rho_v)-\sigma\right\|_1
		\right|
		\le
		\left\|
		\Lambda_t(\rho_u)-\Lambda_t(\rho_v)
		\right\|_1
		=
		\left\|
		\Lambda_t(\rho_u-\rho_v)
		\right\|_1
		\notag\\
		&\le
		\left\|
		\rho_u-\rho_v
		\right\|_1
		=
		2\sqrt{1-|\braket{u}{v}|^2}
		\le
		2\|u-v\|_2 .
		\label{eq:vertical-lipschitz-chain}
	\end{align}
	The inequalities are the reverse triangle inequality, CPTP contractivity, and Lemma~\ref{lem:ranktwo}. Thus \(g_t\) is \(2\)-Lipschitz on the unit sphere. Applying Lemma~\ref{lem:levy} with \(L=2\), and absorbing the factor \(1/4\) into the absolute constant, gives the fixed-time concentration estimate in Theorem~1.

	Assume \(\mu_0>\eps>\lim_{t\to\infty}\mu_t\) and set \(t_*=\tstar(\eps)\). For each \(\psi\), the map \(t\mapsto g_t(\psi)\) is continuous in finite dimension. It is also nonincreasing. Indeed, for \(s\ge t\),
	\begin{align}
		g_s(\psi)
		=
		\left\|
		\Lambda_s(\rho_\psi)-\sigma
		\right\|_1
		=
		\left\|
		\Lambda_{s-t}\!\bigl(\Lambda_t(\rho_\psi)-\sigma\bigr)
		\right\|_1
		\le
		\left\|
		\Lambda_t(\rho_\psi)-\sigma
		\right\|_1
		=
		g_t(\psi).
		\label{eq:vertical-monotone-chain}
	\end{align}
	Since \(0\le g_t(\psi)\le 2\), dominated convergence implies that \(t\mapsto\mu_t\) is continuous and nonincreasing as well. The definition of \(t_*\) then gives \(\mu_{t_*}=\eps\).

	The hitting-time claims are the two one-sided consequences of evaluating the curve at \(t_*\). Monotonicity gives
	\begin{align}
		g_{t_*}(\psi)\le \eps+\eta
		\rightarrow
		\tmix(\rho_\psi,\eps+\eta)\le t_*,
		\qquad
		g_{t_*}(\psi)> \eps-\eta
		\rightarrow
		\tmix(\rho_\psi,\eps-\eta)\ge t_*.
		\label{eq:vertical-hitting-implications}
	\end{align}
	Combining these implications with \(\mu_{t_*}=\eps\) and the fixed-time concentration estimate yields
	\begin{align}
		\Prob_{\mathrm{Haar}}\!\left(
		\tmix(\rho_\psi,\eps+\eta)\le t_*
		\right)
		&\ge
		\Prob_{\mathrm{Haar}}\!\left(
		g_{t_*}(\psi)\le \eps+\eta
		\right)
		=
		1-
		\Prob_{\mathrm{Haar}}\!\left(
		g_{t_*}(\psi)-\mu_{t_*}\ge \eta
		\right)
		\ge
		1-2e^{-cd\eta^2},
		\label{eq:vertical-one-sided-upper}
		\\
		\Prob_{\mathrm{Haar}}\!\left(
		\tmix(\rho_\psi,\eps-\eta)\ge t_*
		\right)
		&\ge
		\Prob_{\mathrm{Haar}}\!\left(
		g_{t_*}(\psi)>\eps-\eta
		\right)
		=
		1-
		\Prob_{\mathrm{Haar}}\!\left(
		\mu_{t_*}-g_{t_*}(\psi)\ge \eta
		\right)
		\ge
		1-2e^{-cd\eta^2}.
		\label{eq:vertical-one-sided-lower}
	\end{align}
	These are the two one-sided hitting-time estimates in Theorem~1.
\end{proof}

\section{Proof of Theorem~2}
\label{app:horizontal}
\begin{proof}[Proof of Theorem~2]
	Write \(\rho_\psi:=\ket{\psi}\!\bra{\psi}\). We test the curve at the two times \(t_*-u\) and \(t_*+u\), where the crossing assumption converts a vertical error into a horizontal one. Fix \(\eta\in(0,m\delta_0]\) and set \(u:=\eta/m\).
	Then $0<u\le \delta_0$, and the transverse-crossing hypothesis gives
	\begin{align}
		\mu_{t_*-u}
		\ge \eps+\eta,\qquad
		\mu_{t_*+u}
		\le \eps-\eta.
		\label{eq:twopointmu}
	\end{align}

	Monotonicity of \(g_t(\psi)\) in \(t\), together with \eqref{eq:twopointmu}, gives the two event inclusions
	\begin{align}
		\left\{
		\tmix(\rho_\psi,\eps)<t_*-u
		\right\}
		\subseteq
		\left\{
		g_{t_*-u}(\psi)\le \eps
		\right\}
		\subseteq
		\left\{
		\mu_{t_*-u}-g_{t_*-u}(\psi)\ge \eta
		\right\},
		\label{eq:horizontal-left-event-chain}
		\\
		\left\{
		\tmix(\rho_\psi,\eps)>t_*+u
		\right\}
		\subseteq
		\left\{
		g_{t_*+u}(\psi)\ge \eps
		\right\}
	\subseteq
		\left\{
		g_{t_*+u}(\psi)-\mu_{t_*+u}\ge \eta
		\right\}.
		\label{eq:horizontal-right-event-chain}
	\end{align}
	Applying Theorem~1 at the two times \(t_*-u\) and \(t_*+u\) yields
	\begin{align}
		\Prob_{\mathrm{Haar}}\!\left(
		\tmix(\rho_\psi,\eps)<t_*-u
		\right)
		\le
		\Prob_{\mathrm{Haar}}\!\left(
		\mu_{t_*-u}-g_{t_*-u}(\psi)\ge \eta
		\right)
		\le
		2e^{-cd\eta^2},
		\label{eq:lefttime}
		\\
		\Prob_{\mathrm{Haar}}\!\left(
		\tmix(\rho_\psi,\eps)>t_*+u
		\right)
		\le
		\Prob_{\mathrm{Haar}}\!\left(
		g_{t_*+u}(\psi)-\mu_{t_*+u}\ge \eta
		\right)
	\le
		2e^{-cd\eta^2}.
		\label{eq:righttime}
	\end{align}
	The union bound now gives
	\begin{align}
		\Prob_{\mathrm{Haar}}\!\left(
		\left|\tmix(\rho_\psi,\eps)-t_*\right|>u
		\right)
		&\le
		\Prob_{\mathrm{Haar}}\!\left(
		\tmix(\rho_\psi,\eps)<t_*-u
		\right)
		+
		\Prob_{\mathrm{Haar}}\!\left(
		\tmix(\rho_\psi,\eps)>t_*+u
		\right)
		\le
		4e^{-cd\eta^2}.
		\label{eq:horizontal-union-chain}
	\end{align}
	Since \(u=\eta/m\), this proves the horizontal concentration bound in Theorem~2.
\end{proof}

\section{Microscopic Many-Body Davies Illustration of Theorem~2}
\label{app:daviesmanybody}

We use a product Davies model for the finite-size check in Fig.~2. The model is intentionally simple. It has a unique product Gibbs state, no rare-state bottleneck, and an exactly factorized finite-time channel \cite{davies1974markovian,kastoryano2016quantum,PhysRevLett.130.060401}. This leaves only the concentration and crossing-time effects to inspect.

For \(n\ge 1\), let
$
	\mcH_n=(\mathbb{C}^2)^{\otimes n},
	\;
	H_n=\frac{\omega}{2}\sum_{j=1}^n (\Id-Z_j),
$
where \(Z_j\) is the Pauli \(Z\) operator on site \(j\), and let
\begin{align}
	\tau_\beta=
	\frac{e^{-\beta \omega |1\rangle\langle 1|}}
	{\Tr(e^{-\beta \omega |1\rangle\langle 1|})}
	=
	\operatorname{diag}(p_0,p_1),
	\qquad
	p_0=\frac{1}{1+e^{-\beta\omega}},
	\qquad
	p_1=\frac{e^{-\beta\omega}}{1+e^{-\beta\omega}}.
	\label{eq:daviesmanybody-tau}
\end{align}
The unique stationary state is the product Gibbs state
$
	\sigma_n=\tau_\beta^{\otimes n}.
$

We thermalize toward this Gibbs state with the local Davies Lindbladian
\begin{align}
	\mcL_n(\rho)=
	\sum_{j=1}^n
	\Bigl(
	\gamma_{\downarrow}\,\mathcal{D}_{F_j^-}(\rho)
	+
	\gamma_{\uparrow}\,\mathcal{D}_{F_j^+}(\rho)
	\Bigr),
	\label{eq:daviesmanybody-L}
\end{align}
where
$
	F_j^-:=|0\rangle\langle 1|_j,
	\;
	F_j^+:=|1\rangle\langle 0|_j,
	\;
	\mathcal{D}_{L}(\rho):=
	L\rho L^\dagger-\frac{1}{2}\{L^\dagger L,\rho\},
$
and \(F_j^\pm\) act on site \(j\) and trivially on the remaining spins. The rates satisfy the detailed-balance relations
$
	\frac{\gamma_{\uparrow}}{\gamma_{\downarrow}}=e^{-\beta\omega},
	\;
	\gamma_{\uparrow}+\gamma_{\downarrow}=1.
$
With this choice, \(\sigma_n\) is the Gibbs fixed point and \(\mcL_n\) is reversible with respect to \(\sigma_n\). Because the local terms in \eqref{eq:daviesmanybody-L} act on distinct tensor factors, they commute. Hence the finite-time channel \(\Lambda_t^{(n)}=e^{t\mcL_n}\) factorizes into identical one-qubit thermal channels. The trace-distance curves
and the full-Haar barycenter benchmark are
\begin{align}
	g_t(\psi)
	:=
	\|\Lambda_t^{(n)}(|\psi\rangle\langle\psi|)-\sigma_n\|_1,\quad
	h_t
	:=
	\|\Lambda_t^{(n)}(\rho_{\mathrm{ref}})-\sigma_n\|_1,
	\quad
	\rho_{\mathrm{ref}}:=\Id/2^n.
	\label{eq:daviesmanybody-curves}
\end{align}
They are evaluated exactly, with no Trotter approximation and no trajectory sampling.

The numerics use
$
	\beta=1.2,
	\;
	\omega=1,
	\;
	\gamma_{\uparrow}+\gamma_{\downarrow}=1,
	\;
	\eps=0.70.
$
We take \(n=3,4,5,6\), so \(d=2^n=8,16,32,64\). For each size, we sample \(48\) Haar-random pure states and evaluate \(g_t(\psi)\) on \(72\) log-spaced times in \([10^{-3},6]\). This range is still small enough for exact computation, but large enough to show the narrowing predicted by Theorem~2.

The left panel of Fig.~2 shows the curve-level picture. The Haar trajectories are already concentrated at fixed time, and the bundle becomes visibly tighter near the common threshold crossing as \(d\) grows. The blue dashed curve is the deterministic barycenter benchmark, which stays below the bulk bundle in accordance with the main-text benchmark inequality. Over the same range, the empirical mean also crosses the threshold with a mildly increasing slope, so the local inverse argument behind Theorem~2 becomes easier to see at larger \(d\).

The right panel of Fig.~2 reports the induced mixing-time statistics. Across \(d=8,16,32,64\), the sample standard deviation of \(t_{\mathrm{mix}}(\psi,\eps)\) drops from about \(0.142\) to \(0.042\), and the \(q_{0.9}-q_{0.1}\) width drops from about \(0.354\) to \(0.106\). The mean crossing stays near \(0.93\)--\(0.95\). The narrowing is therefore not a rescaling of the time axis. It is the finite-size crossing-time version of the fixed-time bundle concentration. The guide \((m_\ast(\eps)\sqrt d)^{-1}\) records the expected conversion from a vertical \(O(d^{-1/2})\) width to a horizontal one.

\section{One-Mode Rare-State Bottleneck Law}
\label{app:rare}
We prove the one-mode estimate used in the main text. The high-probability bound only needs an upper quantile for \(|a_2(\psi)|\). States with smaller slow overlap mix earlier, so they do not enter the bad event.

For Haar-random pure states, define
$
a_2(\psi):=\Tr\!\left(L_2^\dagger \ket{\psi}\!\bra{\psi}\right).
$
For \(\delta\in(0,1)\), define the \((1-\delta)\)-high-probability upper quantile
\begin{equation}
	\alpha_{\mathrm{typ}}^{Q}(\delta)
	:=
	\inf\Bigl\{\alpha\ge 0:\ \Prob_{\mathrm{Haar}}\bigl(|a_2(\psi)|\le \alpha\bigr)\ge 1-\delta\Bigr\}.
	\label{eq:quantile-alpha}
\end{equation}
This is the only probabilistic input needed for the one-mode bottleneck law. No concentration around the mean is assumed at this stage. For later moment-based realizations, we also record
$
	m_2:=\frac{\Tr(L_2)}{d},
	\;
	v_2:=\frac{\Tr(L_2^2)-\Tr(L_2)^2/d}{d(d+1)}.
$

\begin{theorem}[Quantile one-mode rare-state bottleneck law]
	\label{prop:raretail}
	Assume that \(\lambda_2=-\gamma_2\) is real, simple, and strictly dominant, with \(\gamma_2>0\). Fix the corresponding left and right eigenoperators so that \(L_2=L_2^\dagger\), \(R_2=R_2^\dagger\), and \(\|R_2\|_1=1\). Suppose there exist \(\kappa\in(0,1)\) and \(T_0\ge 0\) such that for every pure state \(\ket{\psi}\) and every \(t\ge T_0\),
	\begin{equation}
		\left\|
		\sum_{k\ge 3} e^{\lambda_k t} a_k(\psi) R_k
		\right\|_1
		\le
		\kappa\, |a_2(\psi)|\, e^{-\gamma_2 t}.
		\label{eq:singlemodeassump}
	\end{equation}
	Assume also that the relevant worst-case and \((1-\delta)\)-quantile \(\eps\)-crossings both lie in the one-mode regime \(t\ge T_0\). Then
	\begin{align}
		\tworst(\eps)
		\ge
		\frac{1}{\gamma_2}
		\log\!\left(
		\frac{(1-\kappa)\|L_2\|_\infty}{\eps}
		\right),
		\quad
		t_{\mathrm{typ}}^{(\delta)}(\eps)
		\le
		\frac{1}{\gamma_2}
		\log\!\left(
		\frac{(1+\kappa)\alpha_{\mathrm{typ}}^{Q}(\delta)}{\eps}
		\right).
		\quad
		\Delta t_{\mathrm{tail}}^{(\delta)}(\eps)
		\ge
		\frac{1}{\gamma_2}
		\log\!\left(
		\frac{(1-\kappa)\|L_2\|_\infty}{(1+\kappa)\alpha_{\mathrm{typ}}^{Q}(\delta)}
		\right).
		\label{eq:gaplower}
	\end{align}
\end{theorem}

\begin{proof}
	For a pure state \(\ket{\psi}\), separate the strictly faster tail from the dominant mode by writing
	\begin{align}
		\Delta_t(\psi)
		:=
		\sum_{k\ge 3} e^{\lambda_k t} a_k(\psi) R_k,
		\qquad
		\Lambda_t(\ket{\psi}\!\bra{\psi})-\sigma
		=
		e^{-\gamma_2 t}a_2(\psi)R_2+\Delta_t(\psi).
		\notag
	\end{align}
	Since \(\|R_2\|_1=1\), the single-mode dominance assumption \eqref{eq:singlemodeassump} turns this decomposition into the two-sided trace-norm envelope
	\begin{align}
		g_t(\psi)
		&=
		\left\|
		e^{-\gamma_2 t}a_2(\psi)R_2+\Delta_t(\psi)
		\right\|_1
		\le
		e^{-\gamma_2 t}|a_2(\psi)|+\|\Delta_t(\psi)\|_1
	\le
		(1+\kappa)|a_2(\psi)|e^{-\gamma_2 t},
		\label{eq:uppermode}
		\\
		g_t(\psi)
		&=
		\left\|
		e^{-\gamma_2 t}a_2(\psi)R_2+\Delta_t(\psi)
		\right\|_1
	\ge
		e^{-\gamma_2 t}|a_2(\psi)|-\|\Delta_t(\psi)\|_1
		\ge
		(1-\kappa)|a_2(\psi)|e^{-\gamma_2 t}.
		\label{eq:lowermode}
	\end{align}

	Because \(L_2\) is Hermitian,
	$
	\sup_{\|\psi\|=1}|a_2(\psi)|
	=
	\sup_{\|\psi\|=1}\left|\bra{\psi}L_2\ket{\psi}\right|
	=
	\|L_2\|_\infty.
	$
	Choose a unit vector \(\psi_{\max}\) attaining this supremum. Equation \eqref{eq:lowermode} gives the worst-case lower crossing in
	\begin{align}
		g_t(\psi_{\max})
		\ge
		(1-\kappa)\|L_2\|_\infty e^{-\gamma_2 t},
	\qquad
		t<
		\frac{1}{\gamma_2}
		\log\!\left(\frac{(1-\kappa)\|L_2\|_\infty}{\eps}\right)
		\rightarrow
		g_t(\psi_{\max})>\eps.
		\notag
	\end{align}
	Since the relevant worst-case crossing lies in the one-mode regime, it follows that
	\begin{equation}
		\tworst(\eps)
		\ge
		\frac{1}{\gamma_2}
		\log\!\left(\frac{(1-\kappa)\|L_2\|_\infty}{\eps}\right).
		\label{eq:worstlower}
	\end{equation}

	By definition of \(\alpha_{\mathrm{typ}}^{Q}(\delta)\), the event
	$
	\left\{\psi:\ |a_2(\psi)|\le \alpha_{\mathrm{typ}}^{Q}(\delta)\right\}
	$
	has Haar probability at least \(1-\delta\). On this event, the upper envelope \eqref{eq:uppermode} gives
	\begin{align}
		g_t(\psi)
		\le
		(1+\kappa)\alpha_{\mathrm{typ}}^{Q}(\delta)e^{-\gamma_2 t},
	\qquad
		t\ge
		\frac{1}{\gamma_2}
		\log\!\left(
		\frac{(1+\kappa)\alpha_{\mathrm{typ}}^{Q}(\delta)}{\eps}
		\right)
		\rightarrow
		g_t(\psi)\le \eps.
		\notag
	\end{align}
	Since the relevant \((1-\delta)\)-quantile crossing also lies in the one-mode regime, the definition of \(t_{\mathrm{typ}}^{(\delta)}(\eps)\) yields
	\begin{equation}
		t_{\mathrm{typ}}^{(\delta)}(\eps)
		\le
		\frac{1}{\gamma_2}
		\log\!\left(
		\frac{(1+\kappa)\alpha_{\mathrm{typ}}^{Q}(\delta)}{\eps}
		\right).
		\label{eq:typupper}
	\end{equation}
Subtracting \eqref{eq:typupper} from \eqref{eq:worstlower} gives the last bound in \eqref{eq:gaplower}. Both crossings use the same decay rate \(\gamma_2\). The gap comes from the ratio between the extremal overlap scale \(\|L_2\|_\infty\) and the high-probability upper quantile \(\alpha_{\mathrm{typ}}^{Q}(\delta)\).
\end{proof}

\begin{corollary}[Generic L\'evy realization of the quantile scale]
	\label{cor:raretail-levy}
	There exists an absolute constant \(c_2>0\) such that, for every \(r>0\),
	\begin{equation}
		\Prob_{\mathrm{Haar}}\!\left(
		|a_2(\psi)-m_2|\ge r
		\right)
		\le
		2\exp\!\left(
		-\frac{c_2 d r^2}{\|L_2\|_\infty^2}
		\right).
		\label{eq:a2levy}
	\end{equation}
	Consequently,
	\begin{equation}
		\alpha_{\mathrm{typ}}^{Q}(\delta)
		\le
		|m_2|+\|L_2\|_\infty\sqrt{\frac{\log(2/\delta)}{c_2 d}}.
		\label{eq:quantile-levy}
	\end{equation}
\end{corollary}

\begin{proof}
	For unit vectors \(u\) and \(v\),
	\begin{align*}
		|a_2(u)-a_2(v)|
		=
		\left|
		\Tr\!\left(
		L_2(\ket{u}\!\bra{u}-\ket{v}\!\bra{v})
		\right)
		\right|
		\le
		\|L_2\|_\infty
		\left\|
		\ket{u}\!\bra{u}-\ket{v}\!\bra{v}
		\right\|_1
		\le
		2\|L_2\|_\infty \|u-v\|_2,
	\end{align*}
	where the last step is Lemma~\ref{lem:ranktwo}. Thus \(\psi\mapsto a_2(\psi)\) is \(2\|L_2\|_\infty\)-Lipschitz. L\'evy's lemma, Lemma~\ref{lem:levy}, yields \eqref{eq:a2levy}. If
	$
	r=\|L_2\|_\infty\sqrt{\frac{\log(2/\delta)}{c_2 d}},
	$
	then \eqref{eq:a2levy} gives
	$
	\Prob_{\mathrm{Haar}}\!\left(
		|a_2(\psi)-m_2|\le r
	\right)\ge 1-\delta.
	$
	Hence
	$
	\Prob_{\mathrm{Haar}}\!\left(
		|a_2(\psi)|\le |m_2|+r
	\right)\ge 1-\delta,
	$
	which proves \eqref{eq:quantile-levy}.
\end{proof}

\begin{corollary}[Moment realization of the quantile scale]
	\label{cor:raretail-moment}
	If the second moment \(v_2\) is controlled, then
	\begin{equation}
		\alpha_{\mathrm{typ}}^{Q}(\delta)
		\le
		|m_2|+\sqrt{\frac{v_2}{\delta}}.
		\label{eq:quantile-moment}
	\end{equation}
\end{corollary}

\begin{proof}
	Chebyshev's inequality gives, for any \(r>0\),
	$
		\Prob_{\mathrm{Haar}}\!\left(
		|a_2(\psi)-m_2|\ge r
		\right)
		\le
		\frac{v_2}{r^2}.
		\notag
	$
	Setting \(r=\sqrt{v_2/\delta}\) gives a probability at least \(1-\delta\) event on which
	$
		|a_2(\psi)|
		\le
		|m_2|+|a_2(\psi)-m_2|
		\le
		|m_2|+\sqrt{\frac{v_2}{\delta}}.
		\notag
	$
	The definition \eqref{eq:quantile-alpha} then gives \eqref{eq:quantile-moment}.
\end{proof}

The moment bound is only a convenient sufficient estimate. In the skin branch it is not the useful one, since \(\|L_2\|_2\) may grow with system size while the Haar-typical boundary weight remains small.

\section{Beyond Haar: 2-Design and HS Transfer for Fixed-Time and Slow-Mode Statistics}
\label{app:design}
Full Haar randomness is more than we need for several parts of the argument \cite{dankert2009exact,brandao2016local,PhysRevLett.116.170502,PRXQuantum.4.010311,schuster2025random,zyczkowski2001induced,collins2016random}. Fixed-time concentration can be replaced by an isotropic second-moment estimate. The one-mode bottleneck bound only needs a high-probability upper scale for \(a_2\). We give these two replacements and then apply them to exact \(2\)-designs, induced mixed states, and approximate \(2\)-designs.

For an ensemble \(\mathsf P\) of initial states \(\rho_0\), define
$
	g_t(\rho_0):=\|\Lambda_t(\rho_0)-\sigma\|_1,
	\;
	\mu_t^{(\mathsf P)}:=\E_{\mathsf P}[g_t(\rho_0)],
	\;
	\rho_{\mathrm{ref},\mathsf P}:=\E_{\mathsf P}[\rho_0].
$
The associated deterministic reference curve is
\begin{equation}
	h_t^{(\mathsf P)}:=\|\Lambda_t(\rho_{\mathrm{ref},\mathsf P})-\sigma\|_1.
	\label{eq:ensemble-h}
\end{equation}
The bound \(h_t^{(\mathsf P)}\le \mu_t^{(\mathsf P)}\) is universal. The concentration statement below uses only an isotropic second-moment bound around the maximally mixed barycenter. This is the form needed for the ensembles treated here.

\begin{lemma}[Second-moment transfer for fixed-time and hitting-time concentration]
	\label{lem:fixedtime-transfer}
	For every ensemble \(\mathsf P\) and every \(t\ge 0\),
	\begin{equation}
		h_t^{(\mathsf P)}\le \mu_t^{(\mathsf P)}.
		\label{eq:ensemble-benchmark}
	\end{equation}
	Assume now that \(\rho_{\mathrm{ref},\mathsf P}=\Id/d\). Let \(\{F_a\}_{a=1}^{d^2-1}\) be a traceless Hermitian Hilbert--Schmidt orthonormal basis, and write
	\begin{equation}
		\rho_0-\Id/d=\sum_{a=1}^{d^2-1} x_a(\rho_0)F_a,
		\qquad
		\E_{\mathsf P}[x_a]=0,
		\qquad
		\E_{\mathsf P}[x_a x_b]=c_{\mathsf P}\delta_{ab}.
		\label{eq:ensemble-isotropy}
	\end{equation}
	Define
	$
		\Xi_t^2:=\sum_{a=1}^{d^2-1}\|\Lambda_t(F_a)\|_2^2,
	$
	which is basis independent. Then for every \(\eta>0\),
	\begin{equation}
		\Prob_{\mathsf P}\!\left(
		|g_t(\rho_0)-\mu_t^{(\mathsf P)}|\ge \eta
		\right)
		\le
		\frac{d\,c_{\mathsf P}\,\Xi_t^2}{\eta^2}.
		\label{eq:ensemble-vertical}
	\end{equation}
	If moreover \(\mu_0^{(\mathsf P)}>\eps>\lim_{t\to\infty}\mu_t^{(\mathsf P)}\), \(t_{*,\mathsf P}:=\inf\{t\ge 0:\mu_t^{(\mathsf P)}\le \eps\}\), and
	\begin{equation}
		\mu_{t_{*,\mathsf P}-u}^{(\mathsf P)}\ge \eps+mu,
		\qquad
		\mu_{t_{*,\mathsf P}+u}^{(\mathsf P)}\le \eps-mu,
		\qquad
		0\le u\le \delta_0,
		\label{eq:ensemble-transverse}
	\end{equation}
	then for every \(\eta\in(0,m\delta_0]\), with \(t_\pm:=t_{*,\mathsf P}\pm \eta/m\),
	\begin{equation}
		\Prob_{\mathsf P}\!\left(
		\bigl|\tmix(\rho_0,\eps)-t_{*,\mathsf P}\bigr|>\frac{\eta}{m}
		\right)
		\le
		\frac{d\,c_{\mathsf P}}{\eta^2}
		\bigl(\Xi_{t_-}^2+\Xi_{t_+}^2\bigr).
		\label{eq:ensemble-horizontal}
	\end{equation}
\end{lemma}

\begin{proof}
	
	Equation \eqref{eq:ensemble-benchmark} is Jensen's inequality:
	$
	h_t^{(\mathsf P)}
	=
	\left\|
	\Lambda_t\!\left(\E_{\mathsf P}[\rho_0]\right)-\sigma
	\right\|_1
	\le
	\E_{\mathsf P}\!\left[
	\|\Lambda_t(\rho_0)-\sigma\|_1
	\right]
	=
	\mu_t^{(\mathsf P)}.
	$
	And because \(\mu_t^{(\mathsf P)}\) minimizes the mean squared error among constants,
	\begin{equation}
		\operatorname{Var}_{\mathsf P}(g_t)
		=
		\E_{\mathsf P}\!\left[
		|g_t(\rho_0)-\mu_t^{(\mathsf P)}|^2
		\right]
		\le
		\E_{\mathsf P}\!\left[
		|g_t(\rho_0)-h_t^{(\mathsf P)}|^2
		\right].
		\label{eq:ensemble-varreduce}
	\end{equation}
	Since \(\rho_{\mathrm{ref},\mathsf P}=\Id/d\),
	$
		|g_t(\rho_0)-h_t^{(\mathsf P)}|
		\le
		\left\|
		\Lambda_t\!\left(\rho_0-\frac{\Id}{d}\right)
		\right\|_1
		\le
		\sqrt d\,
		\left\|
		\Lambda_t\!\left(\rho_0-\frac{\Id}{d}\right)
		\right\|_2.
	$
	The centered Hilbert--Schmidt second moment is therefore
	\begin{align}
		\E_{\mathsf P}\!\left[
		\left\|
		\Lambda_t\!\left(\rho_0-\frac{\Id}{d}\right)
		\right\|_2^2
		\right]
		=
		\sum_{a,b=1}^{d^2-1}
		\E_{\mathsf P}[x_a x_b]\,
		\langle \Lambda_t(F_a),\Lambda_t(F_b)\rangle_2
		=
		c_{\mathsf P}\sum_{a=1}^{d^2-1}\|\Lambda_t(F_a)\|_2^2
		=
		c_{\mathsf P}\Xi_t^2.
		\label{eq:ensemble-secondmoment}
	\end{align}
	Combining \eqref{eq:ensemble-varreduce}--\eqref{eq:ensemble-secondmoment} gives
	$
	\operatorname{Var}_{\mathsf P}(g_t)\le d\,c_{\mathsf P}\,\Xi_t^2.
$
	Chebyshev's inequality then yields \eqref{eq:ensemble-vertical}.

	Set \(t_\pm=t_{*,\mathsf P}\pm \eta/m\). Condition \eqref{eq:ensemble-transverse} gives
	$
	\mu_{t_-}^{(\mathsf P)}\ge \eps+\eta,
	\quad
	\mu_{t_+}^{(\mathsf P)}\le \eps-\eta.
	$
	Hence
	\begin{align*}
		\Prob_{\mathsf P}\!\bigl(\tmix(\rho_0,\eps)\le t_-\bigr)
		&\le
		\Prob_{\mathsf P}\!\bigl(g_{t_-}(\rho_0)\le \eps\bigr)
		\le
		\Prob_{\mathsf P}\!\bigl(|g_{t_-}(\rho_0)-\mu_{t_-}^{(\mathsf P)}|\ge \eta\bigr),\\
		\Prob_{\mathsf P}\!\bigl(\tmix(\rho_0,\eps)> t_+\bigr)
		&\le
		\Prob_{\mathsf P}\!\bigl(g_{t_+}(\rho_0)\ge \eps\bigr)
		\le
		\Prob_{\mathsf P}\!\bigl(|g_{t_+}(\rho_0)-\mu_{t_+}^{(\mathsf P)}|\ge \eta\bigr).
	\end{align*}
	Applying \eqref{eq:ensemble-vertical} at \(t_-\) and \(t_+\), then using the union bound, gives \eqref{eq:ensemble-horizontal}. This is the local inverse argument from Theorem~2, with a second-moment estimate replacing L\'evy concentration.
\end{proof}

For the one-mode tail, a moment bound gives one admissible high-probability scale for \(a_2\). For an ensemble \(\mathsf P\) of initial states \(\rho_0\), define
$
	m_{\mathsf P}:=\E_{\mathsf P}[a_2(\rho_0)],
	\;
	v_{\mathsf P}:=\operatorname{Var}_{\mathsf P}(a_2(\rho_0)),
	\;
	\alpha_{\mathsf P}(\delta):=
	|m_{\mathsf P}|+\sqrt{\frac{v_{\mathsf P}}{\delta}},
$
where \(a_2(\rho_0):=\Tr(L_2^\dagger\rho_0)\). We also write
\begin{equation}
	t_{\mathrm{typ},\mathsf P}^{(\delta)}(\eps):=
	\inf\Bigl\{
	t\ge 0:\ 
	\Prob_{\mathsf P}\bigl(\tmix(\rho_0,\eps)\le t\bigr)\ge 1-\delta
	\Bigr\}.
	\label{eq:ensemble-typ}
\end{equation}

\begin{lemma}[Moment realization for the one-mode bottleneck law]
	\label{lem:momenttransfer}
	Assume the hypotheses of Theorem~\ref{prop:raretail}, except that the random initial state is now drawn from an arbitrary ensemble \(\mathsf P\) supported on density matrices \(\rho_0\) satisfying
	\begin{equation}
		\left\|
		\sum_{k\ge 3} e^{\lambda_k t} a_k(\rho_0) R_k
		\right\|_1
		\le
		\kappa\, |a_2(\rho_0)|\, e^{-\gamma_2 t}
		\qquad
		(t\ge T_0)
		\label{eq:ensemble-singlemode}
	\end{equation}
	for every \(\rho_0\) in the support of \(\mathsf P\). If the typical crossing time predicted by the one-mode estimate lies in the regime \(t\ge T_0\), namely if
	$
		\frac{1}{\gamma_2}
		\log\!\left(
		\frac{(1+\kappa)\alpha_{\mathsf P}(\delta)}{\eps}
		\right)
		\ge T_0,
	$
	then
	\begin{equation}
		t_{\mathrm{typ},\mathsf P}^{(\delta)}(\eps)
		\le
		\frac{1}{\gamma_2}
		\log\!\left(
		\frac{(1+\kappa)\alpha_{\mathsf P}(\delta)}{\eps}
		\right).
		\label{eq:ensemble-typupper}
	\end{equation}
	Consequently, if the worst-case lower-bound condition
	$
	\frac{1}{\gamma_2}\log\!\left(\frac{(1-\kappa)\|L_2\|_\infty}{\eps}\right)\ge T_0
	$
	also holds, then
	\begin{equation}
		\tworst(\eps)-t_{\mathrm{typ},\mathsf P}^{(\delta)}(\eps)
		\ge
		\frac{1}{\gamma_2}
		\log\!\left(
		\frac{(1-\kappa)\|L_2\|_\infty}{(1+\kappa)\alpha_{\mathsf P}(\delta)}
		\right).
		\label{eq:ensemble-gap}
	\end{equation}
\end{lemma}

\begin{proof}
	
	Chebyshev's inequality gives
	$
	\Prob_{\mathsf P}\!\left(
	|a_2(\rho_0)-m_{\mathsf P}|
	\ge
	\sqrt{\frac{v_{\mathsf P}}{\delta}}
	\right)
	\le \delta,
	$
	hence
	\begin{equation}
		\Prob_{\mathsf P}\!\left(
		|a_2(\rho_0)|\le \alpha_{\mathsf P}(\delta)
		\right)\ge 1-\delta.
		\label{eq:ensemble-a2event}
	\end{equation}

	Under \eqref{eq:ensemble-singlemode}, the same decomposition used in the proof of Theorem~\ref{prop:raretail} yields, for every state in the support of \(\mathsf P\),
	\begin{equation}
		\|\Lambda_t(\rho_0)-\sigma\|_1
		\le
		(1+\kappa)\,|a_2(\rho_0)|\,e^{-\gamma_2 t},
		\qquad
		(t\ge T_0).
		\label{eq:ensemble-uppermode}
	\end{equation}
	On the event \eqref{eq:ensemble-a2event}, equation \eqref{eq:ensemble-uppermode} implies
$
	\|\Lambda_t(\rho_0)-\sigma\|_1
	\le
	(1+\kappa)\alpha_{\mathsf P}(\delta)e^{-\gamma_2 t},
	\;
	(t\ge T_0).
	$
	Choosing \(t\) as in \eqref{eq:ensemble-typupper} proves \eqref{eq:ensemble-typupper}. Subtracting this from the same worst-case lower bound as in Theorem~\ref{prop:raretail} yields \eqref{eq:ensemble-gap}. Here the moment estimate is used only to produce one admissible upper quantile for the slow overlap.
\end{proof}

We now identify the ensemble constants \(c_{\mathsf P}\) and \(\alpha_{\mathsf P}(\delta)\) for the concrete families used in the main text.

\begin{proposition}[Exact \(2\)-design pure-state transfer]
	\label{prop:exactdesign}
	Let \(\ket{\phi_0}\in\mcH\) be fixed and set
	$
	\rho_U:=U\ket{\phi_0}\!\bra{\phi_0}U^\dagger,
	\quad
	a_2(U):=\Tr(L_2^\dagger\rho_U),
	$
	with \(U\) drawn from an exact unitary \(2\)-design \(\nu\) on \(\mcH\). Then
	\begin{equation}
		\E_\nu[a_2(U)]=m_2,
		\qquad
		\operatorname{Var}_\nu(a_2(U))=v_2,
		\label{eq:exactdesign-moments}
	\end{equation}
 Moreover, if \(\{F_a\}_{a=1}^{d^2-1}\) is a traceless Hermitian Hilbert--Schmidt orthonormal basis and
	$
	x_a(U):=\Tr\!\left[F_a\!\left(\rho_U-\frac{\Id}{d}\right)\right],
	$
	then
	\begin{equation}
		\E_\nu[x_a(U)x_b(U)]
		=
		\frac{\delta_{ab}}{d(d+1)}.
		\label{eq:exactdesign-cov}
	\end{equation}
	Thus Lemma~\ref{lem:fixedtime-transfer} applies with
	$
	c_\nu=\frac{1}{d(d+1)}.
	$
	Hence
	\begin{equation}
		\Prob_\nu\!\left(
		|g_t(\rho_U)-\mu_t^{(\nu)}|\ge \eta
		\right)
		\le
		\frac{\Xi_t^2}{(d+1)\eta^2}
		\qquad
		(t\ge 0),
		\label{eq:exactdesign-vertical}
	\end{equation}
	and, under the same transverse condition \eqref{eq:ensemble-transverse} for \(\mu_t^{(\nu)}\),
	\begin{equation}
		\Prob_\nu\!\left(
		\bigl|\tmix(\rho_U,\eps)-t_{*,\nu}\bigr|>\frac{\eta}{m}
		\right)
		\le
		\frac{\Xi_{t_{-,\nu}}^2+\Xi_{t_{+,\nu}}^2}{(d+1)\eta^2},
		\label{eq:exactdesign-horizontal}
	\end{equation}
	where \(t_{*,\nu}:=\inf\{t:\mu_t^{(\nu)}\le \eps\}\) and \(t_{\pm,\nu}:=t_{*,\nu}\pm \eta/m\). Finally,
	$
	\alpha_\nu(\delta)=|m_2|+\sqrt{\frac{v_2}{\delta}},
	$
	which matches the Haar moment realization from Corollary~\ref{cor:raretail-moment}. Lemma~\ref{lem:momenttransfer} then gives the same one-mode upper bound and gap estimate for the exact \(2\)-design ensemble.
\end{proposition}

\begin{proof}
	Exact \(2\)-designs reproduce the Haar one- and two-fold twirls:
	$
	\E_\nu[\rho_U]=\frac{\Id}{d},
	\quad
	\E_\nu[\rho_U^{\otimes 2}]=\frac{\Id+\mathbb{F}}{d(d+1)}.
	$
	Taking the trace against \(L_2\) and \(L_2\otimes L_2\) gives
	\[
	\E_\nu[a_2(U)]=\frac{\Tr(L_2)}{d}=m_2,
	\qquad
	\E_\nu[|a_2(U)|^2]=\frac{\Tr(L_2)^2+\Tr(L_2^2)}{d(d+1)}.
	\]
	Subtracting the square of the mean yields \(\operatorname{Var}_\nu(a_2(U))=v_2\), which is exactly \eqref{eq:exactdesign-moments}. Repeating the same second-twirl computation with \(F_a\) and \(F_b\) gives
	\[
	\E_\nu[x_a(U)x_b(U)]
	=
	\Tr\!\left[
	(F_a\otimes F_b)\,
	\E_\nu[\rho_U^{\otimes 2}]
	\right]
	=
	\frac{\delta_{ab}}{d(d+1)},
	\]
	because \(\Tr(F_a)=0\) and \(\Tr(F_aF_b)=\delta_{ab}\). This proves \eqref{eq:exactdesign-cov}. The fixed-time and hitting-time bounds then follow immediately from Lemma~\ref{lem:fixedtime-transfer}.
\end{proof}

\begin{proposition}[HS and induced mixed-state transfer]
	\label{prop:inducedtransfer}
	Let \(\mcH_A=\mcH\) with \(\dim\mcH_A=d\), let \(\mcH_B\) have dimension \(d_B\), and fix a pure seed \(\ket{\Phi_0}\in \mcH_A\otimes\mcH_B\). Draw \(U\) from Haar measure, or from an exact unitary \(2\)-design, on \(\mcH_A\otimes\mcH_B\), and define
	\[
	\rho_{AB,U}:=U\ket{\Phi_0}\!\bra{\Phi_0}U^\dagger,
	\qquad
	\rho_{A,U}:=\Tr_B(\rho_{AB,U}),
	\qquad
	a_2(U):=\Tr_A(L_2^\dagger\rho_{A,U}).
	\]
	Then
	\begin{equation}
		\E[a_2(U)]=\frac{\Tr(L_2)}{d}=m_2,
\qquad
		\operatorname{Var}(a_2(U))
		=
		\frac{\Tr(L_2^2)-\Tr(L_2)^2/d}{d(dd_B+1)}.
		\label{eq:induced-var}
	\end{equation}
	Hence
	\begin{equation}
		\alpha_{d_B}(\delta):=
		|m_2|+\sqrt{
			\frac{\Tr(L_2^2)-\Tr(L_2)^2/d}{\delta\, d(dd_B+1)}
		}.
		\label{eq:induced-alpha}
	\end{equation}
	If \(\{F_a\}_{a=1}^{d^2-1}\) is a traceless Hermitian Hilbert--Schmidt orthonormal basis and
	$
	x_a(U):=\Tr\!\left[F_a\!\left(\rho_{A,U}-\frac{\Id}{d}\right)\right],
	$
	then
	\begin{equation}
		\E[x_a(U)x_b(U)]
		=
		\frac{\delta_{ab}}{d(dd_B+1)}.
		\label{eq:induced-cov}
	\end{equation}
	Thus Lemma~\ref{lem:fixedtime-transfer} applies with
	$
	c_{d_B}=\frac{1}{d(dd_B+1)}.
	$
	Hence
	\begin{equation}
		\Prob\!\left(
		|g_t(\rho_{A,U})-\mu_t^{(d_B)}|\ge \eta
		\right)
		\le
		\frac{\Xi_t^2}{(dd_B+1)\eta^2}
		\qquad
		(t\ge 0),
		\label{eq:induced-vertical}
	\end{equation}
	and, under the same transverse condition \eqref{eq:ensemble-transverse} for \(\mu_t^{(d_B)}\),
	\begin{equation}
		\Prob\!\left(
		\bigl|\tmix(\rho_{A,U},\eps)-t_{*,d_B}\bigr|>\frac{\eta}{m}
		\right)
		\le
		\frac{\Xi_{t_{-,d_B}}^2+\Xi_{t_{+,d_B}}^2}{(dd_B+1)\eta^2},
		\label{eq:induced-horizontal}
	\end{equation}
	where \(t_{*,d_B}:=\inf\{t:\mu_t^{(d_B)}\le \eps\}\) and \(t_{\pm,d_B}:=t_{*,d_B}\pm \eta/m\). Finally, Lemma~\ref{lem:momenttransfer} applies with \(\alpha_{\mathsf P}(\delta)=\alpha_{d_B}(\delta)\). In the Hilbert--Schmidt case \(d_B=d\),
	$
		\alpha_{\mathrm{HS}}(\delta)
		:=
		|m_2|+\sqrt{
			\frac{\Tr(L_2^2)-\Tr(L_2)^2/d}{\delta\, d(d^2+1)}
		}.
	$
	If \(\Tr(L_2)=0\), then
	\begin{equation}
		\alpha_{\mathrm{HS}}(\delta)
		=
		\frac{\|L_2\|_2}{\sqrt{\delta\, d(d^2+1)}},
		\label{eq:HS-tracefree}
	\end{equation}
	which is smaller than the pure-state Haar scale by an additional factor \(d^{-1/2}\).
\end{proposition}

\begin{proof}
	
	Set \(D:=dd_B\) and \(X:=L_2\otimes \Id_B\). Then
	$
	a_2(U)=\Tr_{AB}\!\bigl(X\,\rho_{AB,U}\bigr).
	$
	Because Haar measure and exact \(2\)-designs agree on the first two twirls,
	\[
	\E[\rho_{AB,U}]
	=
	\frac{\Id_{AB}}{D},
	\qquad
	\E[\rho_{AB,U}^{\otimes 2}]
	=
	\frac{\Id_{AB}^{\otimes 2}+\mathbb{F}_{AB}}{D(D+1)},
	\]
	where \(\mathbb{F}_{AB}=\mathbb{F}_A\otimes \mathbb{F}_B\). The first identity gives
	$
	\E[a_2(U)]
	=
	\frac{\Tr(X)}{D}
	=
	\frac{d_B\Tr(L_2)}{dd_B}
	=
	\frac{\Tr(L_2)}{d},
	$
For the second moment,
	\begin{align}
		\E[|a_2(U)|^2]
		&=
		\Tr\!\left[
		(X\otimes X)\,
		\E[\rho_{AB,U}^{\otimes 2}]
		\right]
		=
		\frac{\Tr(X)^2+\Tr(X^2)}{D(D+1)}\notag\\
		&=
		\frac{d_B^2\Tr(L_2)^2+d_B\Tr(L_2^2)}{dd_B(dd_B+1)}
		=
		\frac{d_B\Tr(L_2)^2+\Tr(L_2^2)}{d(dd_B+1)}.
		\label{eq:induced-second}
	\end{align}
	Subtracting \(|\E[a_2(U)]|^2=\Tr(L_2)^2/d^2\) yields \eqref{eq:induced-var}.

	The same twirl formula gives
	\begin{equation}
		\E[\rho_{A,U}^{\otimes 2}]
		=
		\frac{d_B\,\Id_A^{\otimes 2}+\mathbb{F}_A}{d(dd_B+1)}.
		\label{eq:induced-rho2}
	\end{equation}
	Tracing \eqref{eq:induced-rho2} against \(F_a\otimes F_b\) yields \eqref{eq:induced-cov}. The fixed-time and hitting-time bounds \eqref{eq:induced-vertical} and \eqref{eq:induced-horizontal} then follow from Lemma~\ref{lem:fixedtime-transfer}. Equations \eqref{eq:induced-alpha}--\eqref{eq:HS-tracefree} are immediate consequences of Lemma~\ref{lem:momenttransfer}. In particular, the ancillary dimension \(d_B\) suppresses the typical overlap by an additional factor \(d_B^{-1/2}\).
\end{proof}

\begin{proposition}[Approximate \(2\)-design transfer]
	\label{prop:approxdesign}
	Consider either of the two exact-design sampling schemes above. One is pure-state sampling on \(\mcH\) as in Proposition~\ref{prop:exactdesign}. The other is induced mixed-state sampling on \(\mcH_A\otimes\mcH_B\) as in Proposition~\ref{prop:inducedtransfer}. Write \(\rho_{\mathrm{sys},U}\) for the sampled state on the system of interest. Thus \(\rho_{\mathrm{sys},U}=\rho_U\) in the pure-state case and \(\rho_{\mathrm{sys},U}=\rho_{A,U}\) in the induced case. Let \(\nu_0\) denote the corresponding exact Haar/exact-design ensemble, with moment parameters \(m_0\) and \(v_0\). Let \(\nu\) be an \(\varepsilon_2\)-approximate unitary \(2\)-design on the relevant unitary space \(\mathcal K\) in diamond norm,
	\begin{equation}
		\bigl\|
		\mathcal T_\nu^{(2)}-\mathcal T_{\nu_0}^{(2)}
		\bigr\|_\diamond
		\le \varepsilon_2.
		\label{eq:approx2design}
	\end{equation}
	Then the slow-overlap moments obey
	\begin{equation}
		\bigl|\E_\nu[a_2]-m_0\bigr|
		\le
		\varepsilon_2\|L_2\|_\infty,
		\label{eq:design-mean-defect}
	\end{equation}
	\begin{equation}
		\bigl|
		\operatorname{Var}_\nu(a_2)-v_0
		\bigr|
		\le
		3\varepsilon_2\|L_2\|_\infty^2.
		\label{eq:designvardefect}
	\end{equation}
	Consequently,
	\begin{equation}
		\alpha_{\nu,\varepsilon_2}(\delta):=
		|m_0|
		+\varepsilon_2\|L_2\|_\infty
		+\sqrt{
			\frac{v_0+3\varepsilon_2\|L_2\|_\infty^2}{\delta}
		}
		\label{eq:alpha-design}
	\end{equation}
	satisfies
	\begin{equation}
		\Prob_\nu\!\left(
		|a_2|
		\le
		\alpha_{\nu,\varepsilon_2}(\delta)
		\right)
		\ge 1-\delta.
		\label{eq:design-typicala2}
	\end{equation}
	Now let \(\bar\rho_\nu:=\E_\nu[\rho_{\mathrm{sys},U}]\), let \(\{F_a\}_{a=1}^{d^2-1}\) be a traceless Hermitian Hilbert--Schmidt orthonormal basis on the system of interest, and define
	$
		\Xi_t^2:=\sum_{a=1}^{d^2-1}\|\Lambda_t(F_a)\|_2^2,
		\;
		\Psi_t^2:=\sum_{a=1}^{d^2-1}\|\Lambda_t^\dagger(F_a)\|_\infty^2.
	$
	If the corresponding exact baseline is the pure-state \(2\)-design ensemble, set \(c_0:=1/[d(d+1)]\). If it is the induced ensemble \(\mathrm{Ind}(d,d_B)\), set \(c_0:=1/[d(dd_B+1)]\). Then
	\begin{equation}
		\E_\nu\!\left[
		\|\Lambda_t(\rho_{\mathrm{sys},U}-\bar\rho_\nu)\|_2^2
		\right]
		\le
		c_0\,\Xi_t^2+3\varepsilon_2\,\Psi_t^2,
		\label{eq:design-fixedtime-second}
	\end{equation}
	and therefore
	\begin{equation}
		\Prob_\nu\!\left(
		|g_t(\rho_{\mathrm{sys},U})-\mu_t^{(\nu)}|\ge \eta
		\right)
		\le
		\frac{d}{\eta^2}\Bigl(c_0\,\Xi_t^2+3\varepsilon_2\,\Psi_t^2\Bigr).
		\label{eq:design-fixedtime-vertical}
	\end{equation}
	If moreover \(\mu_t^{(\nu)}\) obeys the same transverse condition \eqref{eq:ensemble-transverse} around \(t_{*,\nu}:=\inf\{t:\mu_t^{(\nu)}\le \eps\}\), then for every \(\eta\in(0,m\delta_0]\), with \(t_{\pm,\nu}:=t_{*,\nu}\pm \eta/m\),
	\begin{equation}
		\Prob_\nu\!\left(
		\bigl|\tmix(\rho_{\mathrm{sys},U},\eps)-t_{*,\nu}\bigr|>\frac{\eta}{m}
		\right)
		\le
		\frac{d}{\eta^2}
		\sum_{s\in\{t_{-,\nu},t_{+,\nu}\}}
		\Bigl(c_0\,\Xi_s^2+3\varepsilon_2\,\Psi_s^2\Bigr).
		\label{eq:design-fixedtime-horizontal}
	\end{equation}
	Hence Lemma~\ref{lem:momenttransfer}, and therefore the one-mode estimate of Theorem~\ref{prop:raretail}, remains valid after replacing the exact-ensemble scale by \eqref{eq:alpha-design}.
\end{proposition}

\begin{proof}
	
	Write the sampled state on the relevant unitary space \(\mathcal K\) as
	$
	\omega_U:=U\rho_*U^\dagger,
	$
	where \(\rho_*\) is the fixed pure seed for the chosen sampling scheme. In the pure-state case \(\rho_{\mathrm{sys},U}=\omega_U\). In the induced case \(\rho_{\mathrm{sys},U}=\Tr_B(\omega_U)\). Set \(X:=L_2\) in the pure-state case and \(X:=L_2\otimes \Id_B\) in the induced case. Then \(\|X\|_\infty=\|L_2\|_\infty\) and
	$
	a_2=\Tr(X\omega_U),
	\;
	|a_2|^2=\Tr\!\bigl[(X\otimes X)\omega_U^{\otimes 2}\bigr].
	$
	Hence
	\[
	\E_\nu[|a_2|^2]
	=
	\Tr\!\bigl[
	(X\otimes X)\,\mathcal T_\nu^{(2)}(\rho_*^{\otimes 2})
	\bigr],
	\qquad
	\E_{\nu_0}[|a_2|^2]
	=
	\Tr\!\bigl[
	(X\otimes X)\,\mathcal T_{\nu_0}^{(2)}(\rho_*^{\otimes 2})
	\bigr].
	\]
	Therefore,
	\begin{align}
		\bigl|
		\E_\nu[|a_2|^2]-\E_{\nu_0}[|a_2|^2]
		\bigr|
		&=
		\bigl|
		\Tr\!\bigl[
		(X\otimes X)(\mathcal T_\nu^{(2)}-\mathcal T_{\nu_0}^{(2)})(\rho_*^{\otimes 2})
		\bigr]
		\bigr|
		\le
		\|X\otimes X\|_\infty
		\bigl\|
		(\mathcal T_\nu^{(2)}-\mathcal T_{\nu_0}^{(2)})(\rho_*^{\otimes 2})
		\bigr\|_1
		\notag\\
		&\le
		\varepsilon_2\|L_2\|_\infty^2.
		\label{eq:design-second-defect}
	\end{align}
	To control the mean, define \(\mathcal T_\nu^{(1)}(Z):=\E_{U\sim \nu}[UZU^\dagger]\). Fix any state \(\tau\) on a copy of \(\mathcal K\), and define \(\mathcal J(Z):=Z\otimes\tau\). Then
	$
	\mathcal T_\nu^{(1)}
	=
	\Tr_2\circ \mathcal T_\nu^{(2)}\circ \mathcal J,
	$
	and the same factorization holds for \(\nu_0\). Since CPTP maps have diamond norm \(1\),
	\[
	\bigl\|
	\mathcal T_\nu^{(1)}-\mathcal T_{\nu_0}^{(1)}
	\bigr\|_\diamond
	\le
	\bigl\|
	\mathcal T_\nu^{(2)}-\mathcal T_{\nu_0}^{(2)}
	\bigr\|_\diamond
	\le \varepsilon_2.
	\]
	Hence
	\begin{align}
		\bigl|
		\E_\nu[a_2]-\E_{\nu_0}[a_2]
		\bigr|
		=
		\bigl|
		\Tr\!\bigl(
		X[\mathcal T_\nu^{(1)}-\mathcal T_{\nu_0}^{(1)}](\rho_*)
		\bigr)
		\bigr|
		\le
		\|X\|_\infty
		\bigl\|
		[\mathcal T_\nu^{(1)}-\mathcal T_{\nu_0}^{(1)}](\rho_*)
		\bigr\|_1
		\le
		\varepsilon_2\|L_2\|_\infty,
	\end{align}
	which is \eqref{eq:design-mean-defect}. Since
	$
	\bigl|
	|\E_\nu[a_2]|^2-|\E_{\nu_0}[a_2]|^2
	\bigr|
	\le
	2\|L_2\|_\infty\,
	\bigl|
	\E_\nu[a_2]-\E_{\nu_0}[a_2]
	\bigr|,
	$
	equation \eqref{eq:design-mean-defect} gives
	$
	\bigl|
	|\E_\nu[a_2]|^2-|\E_{\nu_0}[a_2]|^2
	\bigr|
	\le
	2\varepsilon_2\|L_2\|_\infty^2.
	$
	Combining this with \eqref{eq:design-second-defect} proves \eqref{eq:designvardefect}. Chebyshev's inequality under \(\nu\) then yields \eqref{eq:design-typicala2}.

	Define
	$
	b_{a,t}(U):=\Tr\!\bigl(\Lambda_t^\dagger(F_a)\rho_{\mathrm{sys},U}\bigr).
	$
	Since \(\Lambda_t(\rho_{\mathrm{sys},U}-\bar\rho_\nu)\) is traceless,
	\begin{equation}
		\E_\nu\!\left[
		\|\Lambda_t(\rho_{\mathrm{sys},U}-\bar\rho_\nu)\|_2^2
		\right]
		=
		\sum_{a=1}^{d^2-1}\operatorname{Var}_\nu(b_{a,t}).
		\label{eq:design-fixedtime-sum}
	\end{equation}
	Applying the same variance-defect estimate \eqref{eq:designvardefect} to each observable on the sampled state gives
	\begin{equation}
		\operatorname{Var}_\nu(b_{a,t})
		\le
		v_{a,t}^{(0)}+3\varepsilon_2\|\Lambda_t^\dagger(F_a)\|_\infty^2,
		\label{eq:design-fixedtime-var}
	\end{equation}
	where \(v_{a,t}^{(0)}\) is the exact-baseline variance. In the induced case, this is the same statement applied to \(X=\Lambda_t^\dagger(F_a)\otimes \Id_B\) on \(\mathcal K=\mcH_A\otimes\mcH_B\). Let
	$
	P_0(X):=X-\frac{\Tr(X)}{d}\Id
	$
	denote the orthogonal projection onto the traceless subspace. Then in the pure-state exact-design baseline,
	$
	v_{a,t}^{(0)}
	=
	\frac{\|P_0\Lambda_t^\dagger(F_a)\|_2^2}{d(d+1)},
	$
	and in the induced baseline \(\mathrm{Ind}(d,d_B)\),
	$
	v_{a,t}^{(0)}
	=
	\frac{\|P_0\Lambda_t^\dagger(F_a)\|_2^2}{d(dd_B+1)}.
	$
	Because \(\Lambda_t\) is trace preserving, it maps the traceless subspace into itself. On that subspace, \(P_0\Lambda_t^\dagger\) is the Hilbert--Schmidt adjoint of \(\Lambda_t\). Hence, for the orthonormal basis \(\{F_a\}_{a=1}^{d^2-1}\),
	\[
	\sum_{a=1}^{d^2-1}\|P_0\Lambda_t^\dagger(F_a)\|_2^2
	=
	\sum_{a=1}^{d^2-1}\|\Lambda_t(F_a)\|_2^2
	=
	\Xi_t^2.
	\]
	Summing \eqref{eq:design-fixedtime-var} over \(a\) and using \eqref{eq:design-fixedtime-sum} proves \eqref{eq:design-fixedtime-second}. The benchmark inequality \(h_t^{(\nu)}\le \mu_t^{(\nu)}\) follows from \eqref{eq:ensemble-benchmark}. Then
	\[
	\operatorname{Var}_\nu(g_t)
	\le
	\E_\nu\!\left[
	|g_t(\rho_{\mathrm{sys},U})-h_t^{(\nu)}|^2
	\right]
	\le
	d\,\E_\nu\!\left[
	\|\Lambda_t(\rho_{\mathrm{sys},U}-\bar\rho_\nu)\|_2^2
	\right].
	\]
	Combining this with \eqref{eq:design-fixedtime-second} and Chebyshev's inequality gives \eqref{eq:design-fixedtime-vertical}. The hitting-time bound \eqref{eq:design-fixedtime-horizontal} follows by repeating the two-time argument from Lemma~\ref{lem:fixedtime-transfer}.
\end{proof}

The use of Haar measure is therefore quite limited. Full Haar sampling gives the exponential fixed-time bound for \(g_t\). Exact \(2\)-designs and induced ensembles retain a weaker fixed-time statement through second moments. Approximate designs add the explicit \(O(\varepsilon_2)\) defects in \eqref{eq:designvardefect} and \eqref{eq:design-fixedtime-second}. The one-mode bottleneck estimate needs even less, namely an upper quantile for \(a_2\).

\section{Separation Hierarchy}

We record the three model estimates quoted in the main text. The skin and boundary examples use the one-mode overlap bound at fixed rate. The protected-sector example uses a different mechanism, since typical states barely enter the slow sector.

\subsection{Skin-Effect Model as a Rare-Tail Application}
\label{app:skin}

Here the one-mode theorem is combined with boundary localization of the left eigenoperator. We use the asymmetric-hopping skin analysis of Ref.~\cite{Ueda_skin_effect} as input and estimate the Haar weight near the boundary.

Consider a one-particle chain of length $L$ with open boundary conditions. In this sector the Hilbert-space dimension is $d=L$. Assume that the leading left eigenoperator $L_2$ is Hermitian, traceless, and exponentially localized near the left boundary in the matrix-element sense
\begin{equation}
	\bigl|
	\bra{x}L_2\ket{y}
	\bigr|
	\le
	\|L_2\|_\infty
	\exp\!\left(
	-\frac{x+y-2}{2\xi}
	\right),
	\qquad
	1\le x,y\le L,
	\label{eq:skinloc}
\end{equation}
for some localization length $\xi>0$ that does not scale with $L$. This is the operator-level version of the left-mode localization proved in Ref.~\cite{Ueda_skin_effect}. We also assume that the single-mode dominance condition \eqref{eq:singlemodeassump} from Theorem~\ref{prop:raretail} holds in the threshold regime. For this application we do not use a variance estimate for \(a_2\). The useful quantity is the weighted boundary mass of a Haar-random state.

\begin{proposition}[Skin-effect logarithmic rare-tail bound]
	\label{prop:skin}
	Under \eqref{eq:skinloc}, for \(\ket{\psi}=\sum_x\psi_x\ket{x}\), set
	\begin{equation}
		\nu_x:=e^{-(x-1)/(2\xi)},
		\qquad
		C_\xi:=\sum_{x=1}^{\infty}\nu_x=\frac{1}{1-e^{-1/(2\xi)}},
		\qquad
		W_\xi(\psi):=\sum_{x=1}^{L}\nu_x|\psi_x|^2.
	\end{equation}
	Then
	\begin{equation}
		|a_2(\psi)|
		\le
		C_\xi\|L_2\|_\infty W_\xi(\psi).
		\label{eq:skinweighted}
	\end{equation}
	Moreover,
	\begin{equation}
		\E_{\mathrm{Haar}}[W_\xi(\psi)]\le \frac{C_\xi}{L},
		\qquad
		\Prob_{\mathrm{Haar}}\!\left(
		W_\xi(\psi)>\frac{C_\xi}{\delta L}
		\right)\le \delta.
		\label{eq:skinmarkov}
	\end{equation}
	Consequently, for every \(\delta\in(0,1)\),
	$
		\alpha_{\mathrm{skin}}(\delta):=
		\frac{C_\xi^2}{\delta L}\|L_2\|_\infty
	$
	is an admissible \((1-\delta)\)-high-probability upper scale for \(|a_2(\psi)|\), and
	\begin{equation}
		\alpha_{\mathrm{typ}}^{Q}(\delta)
		\le
		\alpha_{\mathrm{skin}}(\delta).
		\label{eq:skinalpha}
	\end{equation}
	If in addition the hypotheses of Theorem~\ref{prop:raretail} are satisfied, then
	\begin{equation}
		\tworst(\eps)-t_{\mathrm{typ}}^{(\delta)}(\eps)
		\ge
		\frac{1}{\gamma_2}
		\log\!\left(
		\frac{(1-\kappa)\delta L}{(1+\kappa)C_\xi^2}
		\right).
		\label{eq:skingap}
	\end{equation}
	In particular, for fixed $\xi$, $\delta$, $\kappa$, and $\eps$, the separation grows at least as
	\begin{equation}
		\tworst(\eps)-t_{\mathrm{typ}}^{(\delta)}(\eps)
		\ge
		\frac{\log L}{\gamma_2}+O(1).
		\label{eq:skinlog}
	\end{equation}
\end{proposition}

\begin{proof}
	The localization hypothesis converts the slow overlap into a weighted boundary mass. Indeed, from \eqref{eq:skinloc},
	\begin{align*}
		|a_2(\psi)|
		&=
		\left|
		\sum_{x,y=1}^{L}
		\overline{\psi_x}\bra{x}L_2\ket{y}\psi_y
		\right|
		\le
		\|L_2\|_\infty
		\left(\sum_{x=1}^{L}\nu_x|\psi_x|\right)^2
		\\
		&\le
		\|L_2\|_\infty
		\left(\sum_{x=1}^{L}\nu_x\right)
		\left(\sum_{x=1}^{L}\nu_x|\psi_x|^2\right)
		\le
		C_\xi\|L_2\|_\infty W_\xi(\psi).
	\end{align*}
	The third line is Cauchy--Schwarz, and the last line is exactly \eqref{eq:skinweighted}.

	For Haar-random \(\ket{\psi}\), the coordinates satisfy \(\E_{\mathrm{Haar}}[|\psi_x|^2]=1/L\). Hence
	\begin{align*}
		\E_{\mathrm{Haar}}[W_\xi(\psi)]
		=
		\sum_{x=1}^{L}\nu_x\E_{\mathrm{Haar}}[|\psi_x|^2]
		=
		\frac{1}{L}\sum_{x=1}^{L}\nu_x
		\le
		\frac{C_\xi}{L},
		\qquad
		\Prob_{\mathrm{Haar}}\!\left(
		W_\xi(\psi)>\frac{C_\xi}{\delta L}
		\right)
		\le
		\frac{\E_{\mathrm{Haar}}[W_\xi(\psi)]}{C_\xi/(\delta L)}
		\le
		\delta.
	\end{align*}
	This proves \eqref{eq:skinmarkov}. On the complementary event, the overlap becomes
	$
		|a_2(\psi)|
		\le
		C_\xi\|L_2\|_\infty W_\xi(\psi)
		\le
		C_\xi\|L_2\|_\infty\frac{C_\xi}{\delta L}
		=
		\alpha_{\mathrm{skin}}(\delta).
	$
	Thus \(\alpha_{\mathrm{skin}}(\delta)\) is an admissible high-probability scale and \eqref{eq:skinalpha} follows. Substituting this scale into Theorem~\ref{prop:raretail} gives
	\begin{align*}
		\tworst(\eps)-t_{\mathrm{typ}}^{(\delta)}(\eps)
		\ge
		\frac{1}{\gamma_2}
		\log\!\left(
		\frac{(1-\kappa)\|L_2\|_\infty}
		{(1+\kappa)\alpha_{\mathrm{skin}}(\delta)}
		\right)
		=
		\frac{1}{\gamma_2}
		\log\!\left(
		\frac{(1-\kappa)\delta L}{(1+\kappa)C_\xi^2}
		\right),
	\end{align*}
	which is \eqref{eq:skingap}. Since the logarithm differs from \(\log L\) by an \(L\)-independent constant under the fixed-parameter assumptions, \eqref{eq:skinlog} follows.
\end{proof}

This is the logarithmic branch used in the main text. The worst overlap stays \(O(1)\), while the Haar boundary weight is \(O(L^{-1})\). The resulting gap is \((\log L)/\gamma_2\) up to constants.

\subsection{Boundary-Supported Slow Modes on a Many-Body Chain}
\label{app:boundary}

The many-body version uses the same one-mode estimate, but the dimension behind the typical overlap is now \(q^L\). For this estimate we do not need to specify a detailed boundary Liouvillian. The only input is that the slow left eigenoperator is supported on a fixed boundary block. A concrete unital Pauli-Lindblad realization of this hypothesis is given in Sec.~\ref{app:skinclosingnumerics-manybody}.

Consider an \(L\)-site chain with local dimension \(q\), so the Hilbert-space dimension is \(d=q^L\). Let \(B\) be a boundary block of \(\ell\) sites, with \(\ell=O(1)\) independent of \(L\), and let \(B^c\) denote its complement. Assume that the leading left eigenoperator is Hermitian, traceless, and supported on \(B\).
\begin{equation}
	L_2=A_B\otimes \Id_{B^c},
	\qquad
	A_B=A_B^\dagger,
	\qquad
	\Tr(A_B)=0.
	\label{eq:boundarysupport}
\end{equation}
The assumption means that the slow observable lives on \(O(1)\) degrees of freedom even though the ambient Hilbert space grows exponentially with \(L\). We again assume the single-mode dominance condition \eqref{eq:singlemodeassump}.

\begin{proposition}[Boundary many-body lower bound]
	\label{prop:boundary}
	Under \eqref{eq:boundarysupport},
	\begin{equation}
		\|L_2\|_\infty=\|A_B\|_\infty.
		\label{eq:boundarynorms}
	\end{equation}
	Let \(c_2\) be the absolute constant from Corollary~\ref{cor:raretail-levy}. Then for every \(\delta\in(0,1)\),
	$
		\alpha_{\mathrm{bdry}}(\delta):=
		\|A_B\|_\infty
		\sqrt{\frac{\log(2/\delta)}{c_2 q^L}}
	$
	is an admissible \((1-\delta)\)-high-probability upper scale for \(|a_2(\psi)|\), and
	\begin{equation}
		\alpha_{\mathrm{typ}}^{Q}(\delta)\le \alpha_{\mathrm{bdry}}(\delta).
	\end{equation}
	If in addition the hypotheses of Theorem~\ref{prop:raretail} are satisfied for a threshold \(\eps_L\) whose worst and \((1-\delta)\)-quantile crossings lie in the one-mode regime, then
	\begin{equation}
		\tworst(\eps)-t_{\mathrm{typ}}^{(\delta)}(\eps)
		\ge
		\frac{1}{\gamma_2}
		\log\!\left(
		\frac{(1-\kappa)}{1+\kappa}
		\sqrt{\frac{c_2 q^L}{\log(2/\delta)}}
		\right).
		\label{eq:boundarygap}
	\end{equation}
	Hence, for fixed \(q\), \(\delta\), and \(\kappa\), any such threshold sequence satisfies
	\begin{equation}
		\tworst(\eps)-t_{\mathrm{typ}}^{(\delta)}(\eps)
		\ge
		\frac{L\log q}{2\gamma_2}+O(1).
		\label{eq:boundarylinear}
	\end{equation}
\end{proposition}

\begin{proof}
	The tensor product form in \eqref{eq:boundarysupport} gives the operator norm and removes the Haar mean of the slow overlap through
	\begin{align*}
		\|L_2\|_\infty
		=
		\|A_B\otimes \Id_{B^c}\|_\infty
		=
		\|A_B\|_\infty,
		\qquad
		m_2
		=
		\frac{\Tr(L_2)}{q^L}
		=
		\frac{\Tr(A_B)\Tr(\Id_{B^c})}{q^L}
		=
		0.
	\end{align*}
	The first line proves \eqref{eq:boundarynorms}. Corollary~\ref{cor:raretail-levy}, with \(d=q^L\), then gives
$
		\alpha_{\mathrm{typ}}^{Q}(\delta)
		\le
		|m_2|+\|L_2\|_\infty
		\sqrt{\frac{\log(2/\delta)}{c_2 q^L}}
		=
		\|A_B\|_\infty
		\sqrt{\frac{\log(2/\delta)}{c_2 q^L}}
		=
		\alpha_{\mathrm{bdry}}(\delta).
	$
	Thus \(\alpha_{\mathrm{bdry}}(\delta)\) is an admissible high-probability scale. Inserting it into Theorem~\ref{prop:raretail} yields
	\begin{align*}
		\tworst(\eps)-t_{\mathrm{typ}}^{(\delta)}(\eps)
		\ge
		\frac{1}{\gamma_2}
		\log\!\left(
		\frac{(1-\kappa)\|A_B\|_\infty}
		{(1+\kappa)\alpha_{\mathrm{bdry}}(\delta)}
		\right)
		=
		\frac{1}{\gamma_2}
		\log\!\left(
		\frac{(1-\kappa)}{1+\kappa}
		\sqrt{\frac{c_2 q^L}{\log(2/\delta)}}
		\right).
	\end{align*}
	This proves \eqref{eq:boundarygap}. The fixed-parameter expansion of the logarithm gives \eqref{eq:boundarylinear}.
\end{proof}

Compared with the one-particle skin estimate, the ambient dimension is the only new ingredient. The extremal overlap remains \(O(1)\), while the Haar-typical overlap becomes \(q^{-L/2}\). This produces the linear branch. Boundary-dissipated models may also contain left-right non-orthogonality effects \cite{Mori20resolving,24prb_mori}, which are outside this one-mode estimate.

\subsection{Protected-Sector Rapid-Typical/Slow-Worst Separation}
\label{app:protected}

The protected-sector model gives the exponential branch. This time the rare event is not a large overlap with the same slow mode. It is large initial weight on one protected state. A Haar-random state puts only \(O(d^{-1})\) weight there, so the high-probability scale follows the fast bulk and the worst case follows the slow leakage.

Fix \(d=q^L\), a leakage rate \(\eta_L=e^{-cL}\), and an orthonormal basis \(\{\ket{0}\}\cup\{\ket{i}:i\in B\}\) with \(B=\{1,\dots,d-1\}\). Let
$
	\mcL_L:=\mcL_{\mathrm{deph}}+\mcL_{\mathrm{jump}},
$
where \(\mcL_{\mathrm{deph}}\) dephases in this basis at unit rate, and \(\mcL_{\mathrm{jump}}\) acts on diagonal populations by the classical jump process with rates
$
	r_{0\to i}=r_{i\to 0}=\frac{\eta_L}{d-1},
	\;
	r_{i\to j}=\frac{1}{d-2}
	\;\text{for}\;
	(i\neq j,\ i,j\in B).
$

\begin{proposition}[Typical rapid, worst-case non-rapid]
	\label{prop:protected}
	For this Lindbladian family, the stationary state is
	\(
	\sigma_L=\Id/d
	\).
	Define
	$
		\gamma_S(L):=\eta_L\left(1+\frac{1}{d-1}\right).
	$
	Then, for every \(\eps\in(0,1)\),
	\begin{equation}
		\tworst(\eps)\ge \frac{1}{\gamma_S(L)}\log\!\frac{2(1-1/d)}{\eps}.
		\label{eq:protected-worst}
	\end{equation}
	Moreover, for every \(\delta\in(0,1)\), with Haar probability at least \(1-\delta\),
	\begin{equation}
		g_t(\psi)\le 4e^{-t}
		+2\frac{1+\log(1/\delta)}{d-1}\,e^{-\gamma_S(L)t},
		\qquad (t\ge 0).
		\label{eq:protected-gt}
	\end{equation}
	Consequently, if \(\eps\ge 4(1+\log(1/\delta))/(d-1)\), then
	\begin{equation}
		t_{\mathrm{typ}}^{(\delta)}(\eps)\le \log\!\frac{8}{\eps}.
		\label{eq:protected-typ}
	\end{equation}
	In particular, for every fixed \(\delta\in(0,1)\) and every inverse-polynomial threshold sequence \(\eps_L=L^{-m}\) with fixed \(m>0\), the condition for \eqref{eq:protected-typ} holds for all sufficiently large \(L\), and
	\begin{equation}
		t_{\mathrm{typ}}^{(\delta)}(\eps_L)=O(\log L),
		\qquad
		\tworst(\eps_L)\ge \Omega\bigl(e^{cL}\log L\bigr).
		\label{eq:protected-mismatch}
	\end{equation}
	Hence this family exhibits rapid typical mixing while the conventional worst-case mixing time is non-rapid.
\end{proposition}

\begin{proof}
	Let \(Q_L\) be the classical generator induced by \(\mcL_{\mathrm{jump}}\) on diagonal populations. The diagonal sector contains one protected slow direction and a fast bulk, while the dephasing part removes the off-diagonal sector on an \(O(1)\) time scale. Since the rates are symmetric, the uniform law \(u=d^{-1}(1,\ldots,1)\) is stationary.
	The protected eigenvector is
	$
		b:=e_0-\frac{1}{d-1}\sum_{i\in B}e_i,
	$
	and any population vector decomposes as
	$
		p-u=\left(p_0-\frac{1}{d}\right)b+w(p).
	$
	Here \(w(p)_0=0\) and \(\sum_{i\in B}w(p)_i=0\). Set \(\gamma_B(L)=1+1/(d-2)+\eta_L/(d-1)\). The rates give the diagonal-sector calculation
	\begin{align}
		Q_L e_0
		&=
		-\eta_L e_0+
		\frac{\eta_L}{d-1}\sum_{i\in B}e_i,
		\qquad
		Q_L\!\left(\sum_{i\in B}e_i\right)
		=
		\eta_L e_0-
		\frac{\eta_L}{d-1}\sum_{i\in B}e_i,
		\notag\\
		Q_L b
		=
		Q_L\!\left(
		e_0-\frac{1}{d-1}\sum_{i\in B}e_i
		\right)
		=
		-&\eta_L\left(1+\frac{1}{d-1}\right)b
		=
		-\gamma_S(L)b,
		\qquad
		Q_L w
		=
		-\left(1+\frac{1}{d-2}+\frac{\eta_L}{d-1}\right)w
		=
		-\gamma_B(L)w
		\label{eq:protected-eigs}
	\end{align}
	for every \(w\) with \(w_0=0\) and \(\sum_{i\in B}w_i=0\). Since \(\gamma_B(L)\ge 1\), the population flow and its trace-norm bound are
	\begin{align}
		e^{tQ_L}p-u
		&=
		\left(p_0-\frac{1}{d}\right)e^{-\gamma_S(L)t}b
		+
		e^{-\gamma_B(L)t}w(p).
		\label{eq:protected-popflow}
		\\
		\|e^{tQ_L}p-u\|_1
		&\le
		2\left|p_0-\frac{1}{d}\right|e^{-\gamma_S(L)t}
		+
		e^{-\gamma_B(L)t}\|w(p)\|_1
		\notag\\
		&\le
		2\left|p_0-\frac{1}{d}\right|e^{-\gamma_S(L)t}
		+
		2e^{-t}.
		\label{eq:protected-diagbound}
	\end{align}
	Here \(\|b\|_1=2\) and \(\|w(p)\|_1\le 2(1-p_0)\le 2\).

	Let \(\psi_0:=\ket{0}\). For the initial state \(\ket{0}\!\bra{0}\), one has \(w=0\), so \eqref{eq:protected-popflow} is exact. Hence
	\begin{align}
		e^{t\mcL_L}(\ket{0}\!\bra{0})-\sigma_L
		=
		\left(1-\frac{1}{d}\right)e^{-\gamma_S(L)t}
		\left(
		\ket{0}\!\bra{0}
		-
		\frac{1}{d-1}\sum_{i\in B}\ket{i}\!\bra{i}
		\right).
	\qquad
		g_t(\psi_0)
		=
		2\left(1-\frac{1}{d}\right)e^{-\gamma_S(L)t},
		\label{eq:protected-zero-distance}
	\end{align}
	which implies \eqref{eq:protected-worst}.

	Let \(P_x:=\ket{x}\!\bra{x}\) and define the diagonal pinching map
	$
		\Delta(\rho):=\sum_{x=0}^{d-1}P_x\rho P_x.
	$
	The generators \(\mcL_{\mathrm{deph}}\) and \(\mcL_{\mathrm{jump}}\) commute. Moreover,
	$
		e^{t\mcL_{\mathrm{deph}}}(\rho)
		=
		\Delta(\rho)+e^{-t}\bigl(\rho-\Delta(\rho)\bigr).
	$
	Together with trace-norm contractivity of \(e^{t\mcL_{\mathrm{jump}}}\), this gives
	\begin{align}
		e^{t\mcL_L}(\rho)
		=
		e^{t\mcL_{\mathrm{jump}}}\!\left(\Delta(\rho)+e^{-t}\bigl(\rho-\Delta(\rho)\bigr)\right),
		\qquad
		\left\|
		e^{t\mcL_L}(\rho)-e^{t\mcL_{\mathrm{jump}}}(\Delta(\rho))
		\right\|_1
		\le
		e^{-t}\|\rho-\Delta(\rho)\|_1
		\le
		2e^{-t}.
		\label{eq:protected-offdiag}
	\end{align}

	Now take \(\rho=\ket{\psi}\!\bra{\psi}\) and set \(p_0(\psi):=|\langle 0|\psi\rangle|^2\).
	Combining \eqref{eq:protected-diagbound} and \eqref{eq:protected-offdiag} gives
	\begin{equation}
		g_t(\psi)
		\le
		4e^{-t}
		+
		2\left|p_0(\psi)-\frac{1}{d}\right|e^{-\gamma_S(L)t}.
		\label{eq:protected-masterbound}
	\end{equation}
	For Haar-random \(\ket{\psi}\), the variable \(p_0(\psi)\) has the \(\mathrm{Beta}(1,d-1)\) law, so
	\begin{align}
		\Prob_{\mathrm{Haar}}\!\bigl(p_0(\psi)\ge s\bigr)
		=
		\int_s^1 (d-1)(1-p)^{d-2}\,dp
		=
		(1-s)^{d-1}
		\le e^{-(d-1)s}.
		\label{eq:protected-beta}
	\end{align}
	Choosing \(s=\log(1/\delta)/(d-1)\) yields an event \(E_\delta\) of probability at least \(1-\delta\) on which
	\begin{align}
		\left|p_0(\psi)-\frac{1}{d}\right|
		\le
		\frac{\log(1/\delta)}{d-1}+\frac{1}{d}
		\le
		\frac{\log(1/\delta)}{d-1}+\frac{1}{d-1}
		=
		\frac{1+\log(1/\delta)}{d-1}.
		\label{eq:protected-p0}
	\end{align}
	Substituting \eqref{eq:protected-p0} into \eqref{eq:protected-masterbound} proves \eqref{eq:protected-gt}.
	
	Assume \(\eps\ge 4(1+\log(1/\delta))/(d-1)\) and \(t\ge \log(8/\eps)\). Then \eqref{eq:protected-gt} gives
	\begin{align*}
		g_t(\psi)
		\le
		4e^{-t}
		+
		2\frac{1+\log(1/\delta)}{d-1}\,e^{-\gamma_S(L)t}
		\le
		\frac{\eps}{2}
		+
		2\frac{1+\log(1/\delta)}{d-1}
		\le
		\eps
	\end{align*}
	on \(E_\delta\). This proves \eqref{eq:protected-typ}. For \(\eps_L=L^{-m}\), the threshold condition holds for all sufficiently large \(L\) because \(d=q^L\) grows exponentially. Moreover,
	\begin{align*}
		\gamma_S(L)^{-1}
		=
		\frac{e^{cL}}{1+1/(d-1)}
		=
		e^{cL}(1+o(1)),
		\qquad
		\tworst(\eps_L)
		\ge
		\frac{1}{\gamma_S(L)}
		\log\!\frac{2(1-1/d)}{\eps_L}
		=
		\Omega(e^{cL}\log L).
	\end{align*}
	This proves \eqref{eq:protected-mismatch}.
\end{proof}

This construction separates the two mechanisms. In the boundary example, typical and worst states probe the same slow mode with different overlaps. Here the slow time belongs to a sector that a Haar state almost never occupies. For inverse-polynomial accuracies, the quantile benchmark need not resolve that sector. We next write the same idea in logical-sector language \cite{25arxiv_rapid_information}.

\subsection{Logical-Sector Product Model Inspired by Self-Correcting Memories}
\label{app:logicalsector}

We now put the protected-sector idea into a logical-syndrome tensor product. The model is still exactly solvable, but its structure is closer to the logical-sector picture used in self-correcting-memory discussions. The logical factor carries the leakage rate. The syndrome factor relaxes quickly.

The starting point is the logical-syndrome factorization
\begin{equation*}
	\rho_\beta
	\cong
	\frac{\Id_{\mathrm{log}}}{D_{\mathrm{log}}}\otimes \pi_\beta,
	\qquad
	\rho_\beta^\psi
	\cong
	\psi\otimes \pi_\beta,
\end{equation*}
This is meant as a trace-distance version of the two-timescale picture in Ref.~\cite{25arxiv_rapid_information}. Random ground-state initialization there becomes mixed over the logical label only after logical leakage. Here we use full-Haar pure states on the logical-syndrome Hilbert space so that the typical benchmark is defined relative to one fixed target state.

\begin{figure}[t]
	\centering
	\includegraphics[width=\textwidth]{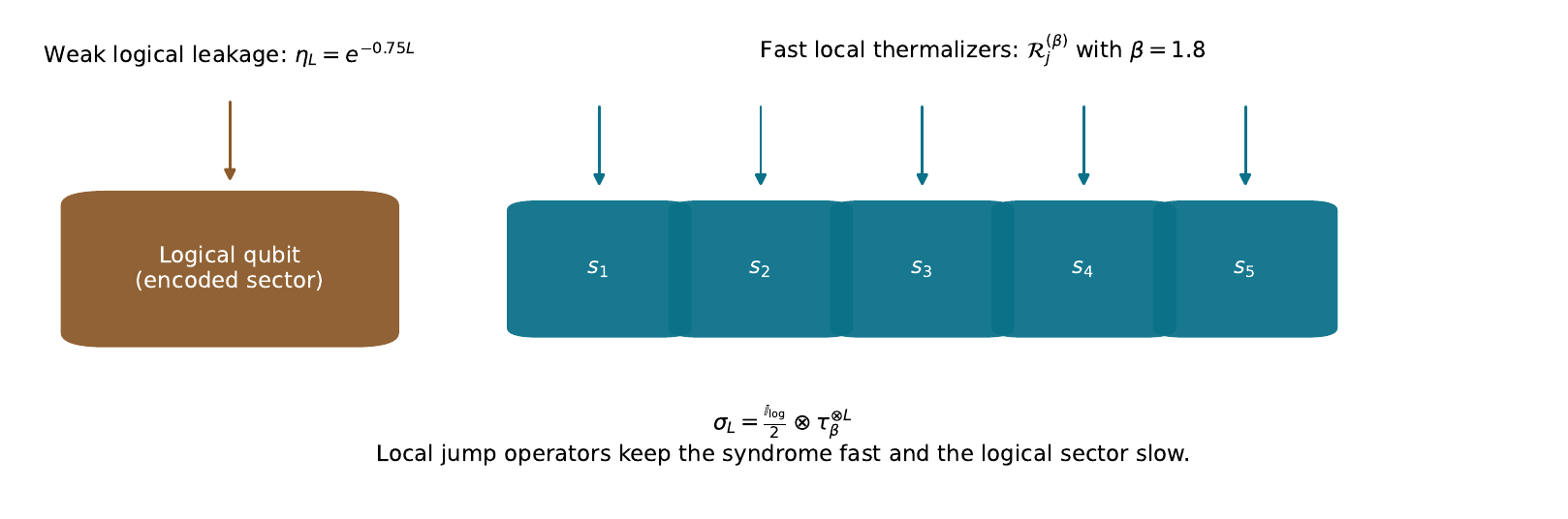}
	\caption{\textbf{Microscopic local-qubit refinement of the logical-sector product model.} The refinement \eqref{eq:logicalmicro-L} specializes the logical factor to one weakly depolarized qubit with leakage rate \(\eta_L=e^{-cL}\), while the syndrome factor is a register of rapidly reset qubits with one-qubit Gibbs state \(\tau_\beta\). The stationary state is \(\sigma_L=(\Id_{\mathrm{log}}/2)\otimes \tau_\beta^{\otimes L}\).}
	\label{fig:logicalschematic}
\end{figure}

\begin{figure}[t]
	\centering
	\includegraphics[width=0.8\textwidth]{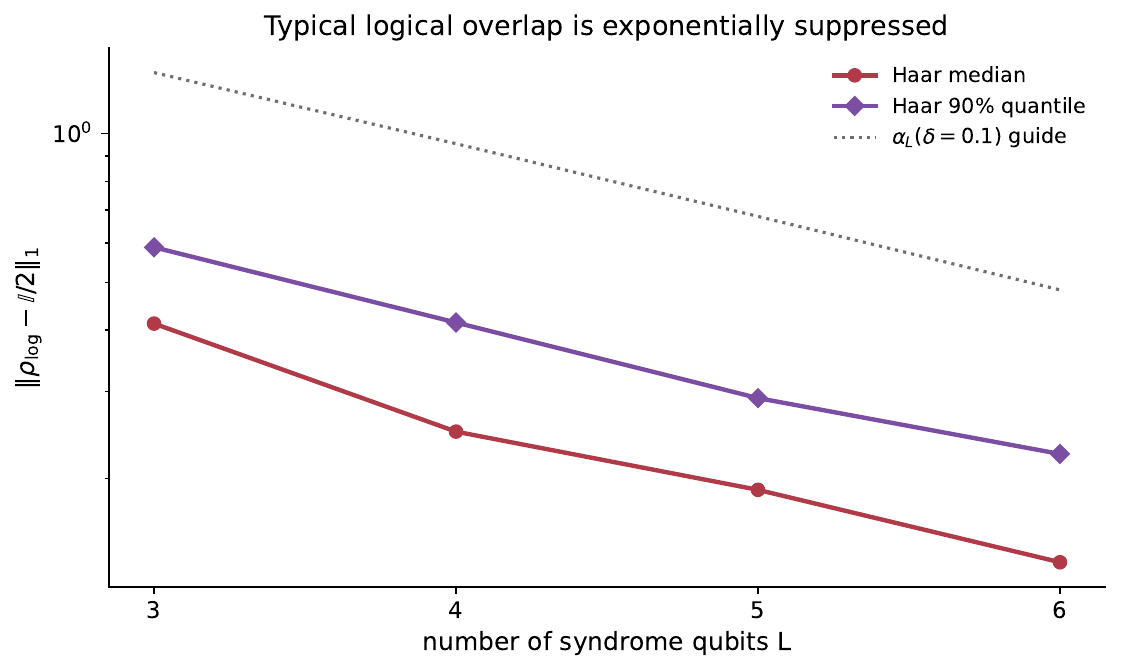}
	\caption{\textbf{Logical-overlap suppression for Haar-random initial states in the local-qubit refinement.} The median and \(90\%\) quantile of \(\Delta_{\mathrm{log}}(\psi)\) decrease with \(L\). The curve \eqref{eq:logicalmicro-alpha} at \(\delta=0.1\) gives the corresponding high-probability bound. Generic initial states place only a small logical mismatch on the slow sector, even though a logical-basis state remains slow.}
	\label{fig:logicaloverlap}
\end{figure}

\begin{proposition}[Logical-sector product model]
	\label{prop:logicalsector}
	Fix an integer \(D\ge 2\) and a sequence \(N_L\to\infty\). Let
	\begin{equation}
		\mcH_L=\mathbb{C}^{D}\otimes\mathbb{C}^{N_L},
		\qquad
		\sigma_L:=\frac{\Id_D}{D}\otimes \pi_L,
		\label{eq:logicalsector-sigma}
	\end{equation}
	where \(\pi_L\) is any state on \(\mathbb{C}^{N_L}\). Define the logical depolarizing and syndrome-reset generators by
	\begin{equation}
		\mathcal{D}_{\mathrm{log}}(X):=\frac{\Tr(X)}{D}\Id_D-X,
		\qquad
		\mathcal{R}_L(Y):=\Tr(Y)\pi_L-Y,
		\label{eq:logicalsector-generators}
	\end{equation}
	and fix a leakage rate \(\eta_L\in(0,1]\). On \(\mcH_L\), consider the Lindbladian
	\begin{equation}
		\mcL_L
		:=
		\eta_L\,\mathcal{D}_{\mathrm{log}}\otimes \mathrm{id}_{\mathrm{syn}}
		+
		\mathrm{id}_{\mathrm{log}}\otimes \mathcal{R}_L.
		\label{eq:logicalsector-L}
	\end{equation}
	Write \(g_t(\psi)\), \(t_{\mathrm{typ}}^{(\delta)}(\eps)\), and \(\tworst(\eps)\) for this model relative to the fixed point \(\sigma_L\). For \(\delta\in(0,1)\), define
	\begin{equation}
		\alpha_L(\delta)
		:=
		\sqrt{\frac{D^2-1}{\delta(DN_L+1)}}.
		\label{eq:logicalsector-alpha}
	\end{equation}
	Then \(\sigma_L\) is the unique stationary state. For every \(\eps\in(0,2(1-1/D))\),
	\begin{equation}
		\tworst(\eps)
		\ge
		\frac{1}{\eta_L}
		\log\!\frac{2(1-1/D)}{\eps}.
		\label{eq:logicalsector-worst}
	\end{equation}
	For Haar-random pure states on \(\mcH_L\), with probability at least \(1-\delta\),
	\begin{equation}
		g_t(\psi)
		\le
		\alpha_L(\delta)e^{-\eta_L t}+6e^{-t}
		\qquad
		(t\ge 0).
		\label{eq:logicalsector-gt}
	\end{equation}
	If
	\begin{equation}
		\eps\ge 2\alpha_L(\delta),
		\label{eq:logicalsector-threshold}
	\end{equation}
	then
	\begin{equation}
		t_{\mathrm{typ}}^{(\delta)}(\eps)
		\le
		\log\!\frac{12}{\eps}.
		\label{eq:logicalsector-typ}
	\end{equation}
	Consequently,
	\begin{equation}
		\tworst(\eps)-t_{\mathrm{typ}}^{(\delta)}(\eps)
		\ge
		\frac{1}{\eta_L}
		\log\!\frac{2(1-1/D)}{\eps}
		-
		\log\!\frac{12}{\eps}.
		\label{eq:logicalsector-gap}
	\end{equation}
	In particular, if \(D\) is fixed, \(N_L\) grows exponentially with system size, and \(\eta_L^{-1}\) grows superlogarithmically, then inverse-polynomial thresholds satisfy \(t_{\mathrm{typ}}^{(\delta)}(\eps_L)=O(\log \eps_L^{-1})\), while the worst-case mixing time is controlled by the much slower leakage scale \(\eta_L^{-1}\).
\end{proposition}

\begin{proof}
	
	Since \(\mathcal{D}_{\mathrm{log}}\) fixes multiples of \(\Id_D\) and acts as \(-\mathrm{id}\) on traceless operators,
	\begin{equation}
		e^{s\mathcal{D}_{\mathrm{log}}}(X)
		=
		e^{-s}X+(1-e^{-s})\frac{\Tr(X)}{D}\Id_D,
		\qquad
		(s\ge 0).
		\label{eq:logicalsector-depol}
	\end{equation}
	Likewise, \(\mathcal{R}_L\) fixes multiples of \(\pi_L\) and acts as \(-\mathrm{id}\) on the kernel of the trace, so
	\begin{equation}
		e^{t\mathcal{R}_L}(Y)
		=
		e^{-t}Y+(1-e^{-t})\Tr(Y)\pi_L,
		\qquad
		(t\ge 0).
		\label{eq:logicalsector-reset}
	\end{equation}
	The two generators commute because they act on different tensor factors. Hence
	\begin{equation}
		e^{t\mcL_L}
		=
		e^{\eta_L t\mathcal{D}_{\mathrm{log}}}\otimes e^{t\mathcal{R}_L},
		\label{eq:logicalsector-semigroup}
	\end{equation}
	and \(\sigma_L\) from \eqref{eq:logicalsector-sigma} is the unique stationary state.

	Fix any unit vector \(\phi\in\mathbb{C}^D\) and set
	$
		\rho_0=\ket{\phi}\!\bra{\phi}\otimes \pi_L.
	$
	Because the syndrome factor is already stationary, the slow logical trajectory is
	\begin{align}
		e^{t\mcL_L}(\rho_0)-\sigma_L
		&=
		e^{-\eta_L t}
		\left(
		\ket{\phi}\!\bra{\phi}-\frac{\Id_D}{D}
		\right)\otimes \pi_L.
		\label{eq:logicalsector-exacttraj}
		\\
		\left\|
		e^{t\mcL_L}(\rho_0)-\sigma_L
		\right\|_1
		&=
		e^{-\eta_L t}
		\left\|
		\ket{\phi}\!\bra{\phi}-\frac{\Id_D}{D}
		\right\|_1
		=
		2\left(1-\frac{1}{D}\right)e^{-\eta_L t}.
		\label{eq:logicalsector-exactslow}
	\end{align}
	The last line uses that \(\ket{\phi}\!\bra{\phi}-\Id_D/D\) has eigenvalue \(1-1/D\) once and eigenvalue \(-1/D\) with multiplicity \(D-1\).
	To lift this mixed-state trajectory to the pure-state worst case, write a spectral decomposition
	\begin{equation*}
		\pi_L=\sum_j p_j \ket{j}\!\bra{j},
		\qquad
		\ket{\Psi_j}:=\ket{\phi}\otimes \ket{j}.
	\end{equation*}
	For each \(j\), set \(g_t(\Psi_j)=\left\|
	e^{t\mcL_L}(\ket{\Psi_j}\!\bra{\Psi_j})-\sigma_L
	\right\|_1\). 
	Then \(\rho_0=\sum_j p_j \ket{\Psi_j}\!\bra{\Psi_j}\), so convexity of the trace norm gives
	\begin{align}
		2\left(1-\frac{1}{D}\right)e^{-\eta_L t}
		=
		\left\|
		e^{t\mcL_L}(\rho_0)-\sigma_L
		\right\|_1
		\le
		\sum_j p_j\,g_t(\Psi_j)
		\le
		\sup_j g_t(\Psi_j).
		\label{eq:logicalsector-convex}
	\end{align}
	Hence for every \(t<\eta_L^{-1}\log(2(1-1/D)/\eps)\), some \(j\) satisfies \(g_t(\Psi_j)>\eps\). Taking the supremum over such \(t\) proves \eqref{eq:logicalsector-worst}.

	Let \(\rho_\psi:=\ket{\psi}\!\bra{\psi}\), and define the reduced states
$
		\rho_{\mathrm{log}}(\psi):=\Tr_{\mathrm{syn}}(\rho_\psi),
		\quad
		\rho_{\mathrm{syn}}(\psi):=\Tr_{\mathrm{log}}(\rho_\psi).
$
	Introduce the correlation remainder
	\begin{equation}
		C(\psi)
		:=
		\rho_\psi
		-
		\rho_{\mathrm{log}}(\psi)\otimes \pi_L
		-
		\frac{\Id_D}{D}\otimes \rho_{\mathrm{syn}}(\psi)
		+
		\sigma_L.
		\label{eq:logicalsector-C}
	\end{equation}
	Using \eqref{eq:logicalsector-semigroup}, \eqref{eq:logicalsector-depol}, and \eqref{eq:logicalsector-reset}, we obtain
	\begin{align}
		e^{t\mcL_L}(\rho_\psi)-\sigma_L
		=
		e^{-\eta_L t}
		\left(
		\rho_{\mathrm{log}}(\psi)-\frac{\Id_D}{D}
		\right)\otimes \pi_L
		+e^{-t}\frac{\Id_D}{D}\otimes
		\left(
		\rho_{\mathrm{syn}}(\psi)-\pi_L
		\right)
		+e^{-(1+\eta_L)t}C(\psi).
		\label{eq:logicalsector-decomp}
	\end{align}
	Let
	$
		\Delta_{\mathrm{log}}(\psi)
		:=
		\left\|
		\rho_{\mathrm{log}}(\psi)-\frac{\Id_D}{D}
		\right\|_1.
	$
	Because \(\|\rho_{\mathrm{syn}}(\psi)-\pi_L\|_1\le 2\), \(\|C(\psi)\|_1\le 4\), and \(\eta_L\le 1\), equation \eqref{eq:logicalsector-decomp} yields
	\begin{equation}
		g_t(\psi)
		\le
		e^{-\eta_L t}\Delta_{\mathrm{log}}(\psi)
		+2e^{-t}
		+4e^{-(1+\eta_L)t}
		\le
		e^{-\eta_L t}\Delta_{\mathrm{log}}(\psi)+6e^{-t}.
		\label{eq:logicalsector-basicbound}
	\end{equation}
	It remains to control \(\Delta_{\mathrm{log}}(\psi)\) under Haar sampling.
	
	Let \(F_{\mathrm{log}}\) and \(F_{\mathrm{syn}}\) denote the swaps on \((\mathbb{C}^{D})^{\otimes 2}\) and \((\mathbb{C}^{N_L})^{\otimes 2}\), respectively. By the same second-twirl identity used in Lemma~\ref{lem:haarmoments},
	$
		\E_{\mathrm{Haar}}\!\left[\rho_\psi^{\otimes 2}\right]
		=
		\frac{\Id+F_{\mathrm{log}}\otimes F_{\mathrm{syn}}}{DN_L(DN_L+1)}.
	$
	Using the swap trick for the purity of the logical marginal,
	\begin{align}
		\E_{\mathrm{Haar}}\!\left[\Tr\!\left(\rho_{\mathrm{log}}(\psi)^2\right)\right]
		&=
		\Tr\!\left[
		\bigl(F_{\mathrm{log}}\otimes \Id_{\mathrm{syn}\,\mathrm{syn}'}\bigr)
		\E_{\mathrm{Haar}}\!\left[\rho_\psi^{\otimes 2}\right]
		\right]
		\notag\\
		&=
		\frac{
			\Tr(F_{\mathrm{log}})\Tr(\Id_{\mathrm{syn}\,\mathrm{syn}'})
			+
			\Tr(\Id_{\mathrm{log}\,\mathrm{log}'})\Tr(F_{\mathrm{syn}})
		}{
			DN_L(DN_L+1)
		}
		\notag\\
		&=
		\frac{DN_L^2+D^2N_L}{DN_L(DN_L+1)}
		=
		\frac{D+N_L}{DN_L+1}.
		\label{eq:logicalsector-purity}
	\end{align}
	The trace-to-Hilbert--Schmidt comparison and \eqref{eq:logicalsector-purity} give
	\begin{align}
		\Delta_{\mathrm{log}}(\psi)^2
		&\le
		D
		\left\|
		\rho_{\mathrm{log}}(\psi)-\frac{\Id_D}{D}
		\right\|_2^2
		=
		D\left(
		\Tr\!\left(\rho_{\mathrm{log}}(\psi)^2\right)-\frac{1}{D}
		\right),
		\notag\\
		\E_{\mathrm{Haar}}\!\left[\Delta_{\mathrm{log}}(\psi)^2\right]
		&\le
		D\left(
		\frac{D+N_L}{DN_L+1}-\frac{1}{D}
		\right)
		=
		\frac{D^2-1}{DN_L+1}.
		\label{eq:logicalsector-secondmoment}
	\end{align}
	Markov's inequality therefore yields
	\begin{align*}
		\Prob_{\mathrm{Haar}}\!\left(
		\Delta_{\mathrm{log}}(\psi)\ge \alpha_L(\delta)
		\right)
		&\le
		\frac{
		\E_{\mathrm{Haar}}\!\left[\Delta_{\mathrm{log}}(\psi)^2\right]
		}{\alpha_L(\delta)^2}
		\le
		\delta.
	\end{align*}
	On the complementary event, \eqref{eq:logicalsector-basicbound} gives
	\begin{align*}
		g_t(\psi)
		&\le
		\alpha_L(\delta)e^{-\eta_L t}+6e^{-t},
		\qquad
		(t\ge 0),
	\end{align*}
	which proves \eqref{eq:logicalsector-gt}.

	Assume now that \eqref{eq:logicalsector-threshold} holds and \(t\ge \log(12/\eps)\). Then \eqref{eq:logicalsector-gt} implies
	\begin{align*}
		g_t(\psi)
		\le
		\alpha_L(\delta)e^{-\eta_L t}+6e^{-t}
		\le
		\frac{\eps}{2}+\frac{\eps}{2}
		=
		\eps
	\end{align*}
	with Haar probability at least \(1-\delta\). This proves \eqref{eq:logicalsector-typ}. Subtracting \eqref{eq:logicalsector-typ} from \eqref{eq:logicalsector-worst} yields \eqref{eq:logicalsector-gap}.
\end{proof}

This product model is the logical-sector version of Sec.~\ref{app:protected}. The fast part resets the syndrome, while the slow part acts only on the logical factor. A Haar-random state is nearly maximally mixed on the fixed logical subsystem, so the slow logical contribution is typically of size \(\alpha_L(\delta)\). The worst case is still a logical state and therefore follows the leakage time.

\begin{figure*}[t]
	\centering
	\includegraphics[width=0.90\textwidth]{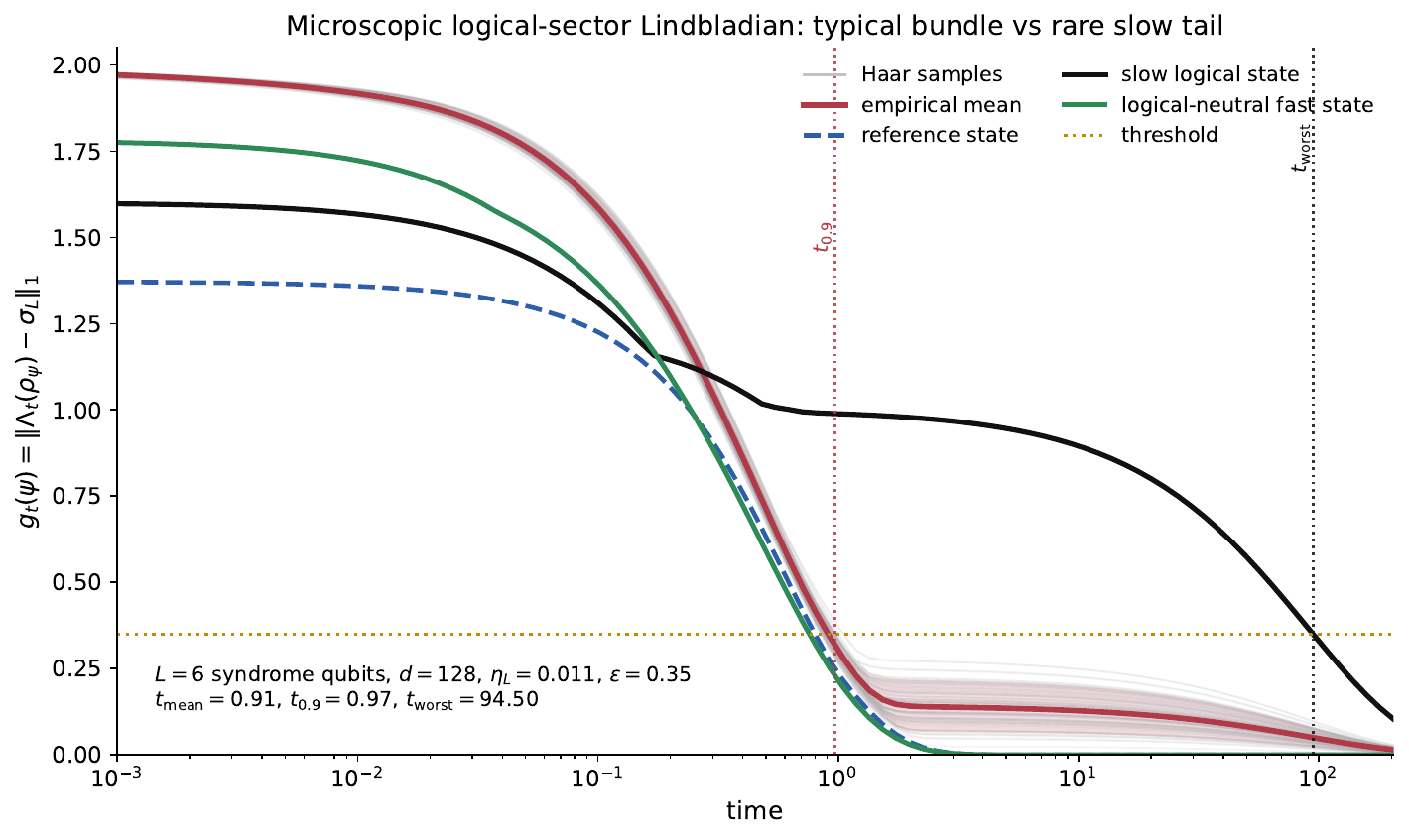}
	\caption{\textbf{Microscopic local-qubit realization of the logical-sector bundle.} Exact relaxation curves for the local model \eqref{eq:logicalmicro-L} at \(L=6\), \(d=2^{L+1}=128\), and threshold \(\eps=0.35\). Gray curves are Haar-random pure-state trajectories \(g_t(\psi)\). The red curve is the empirical mean \(\mu_t\). The blue dashed curve is the reference trajectory generated by \(\rho_{\mathrm{ref}}=\Id/d\). The black curve is the slow benchmark \(\rho_{\mathrm{slow}}\) from \eqref{eq:logicalmicro-slow}. The green curve is the fast benchmark \(\rho_{\mathrm{fast}}=|\psi_{\mathrm{fast}}\rangle\langle\psi_{\mathrm{fast}}|\) from \eqref{eq:logicalmicro-fast}. The product-model mechanism persists in this strictly local qubit realization.}
	\label{fig:logicalbundle}
\end{figure*}

\begin{figure*}[t]
	\centering
	\includegraphics[width=0.90\textwidth]{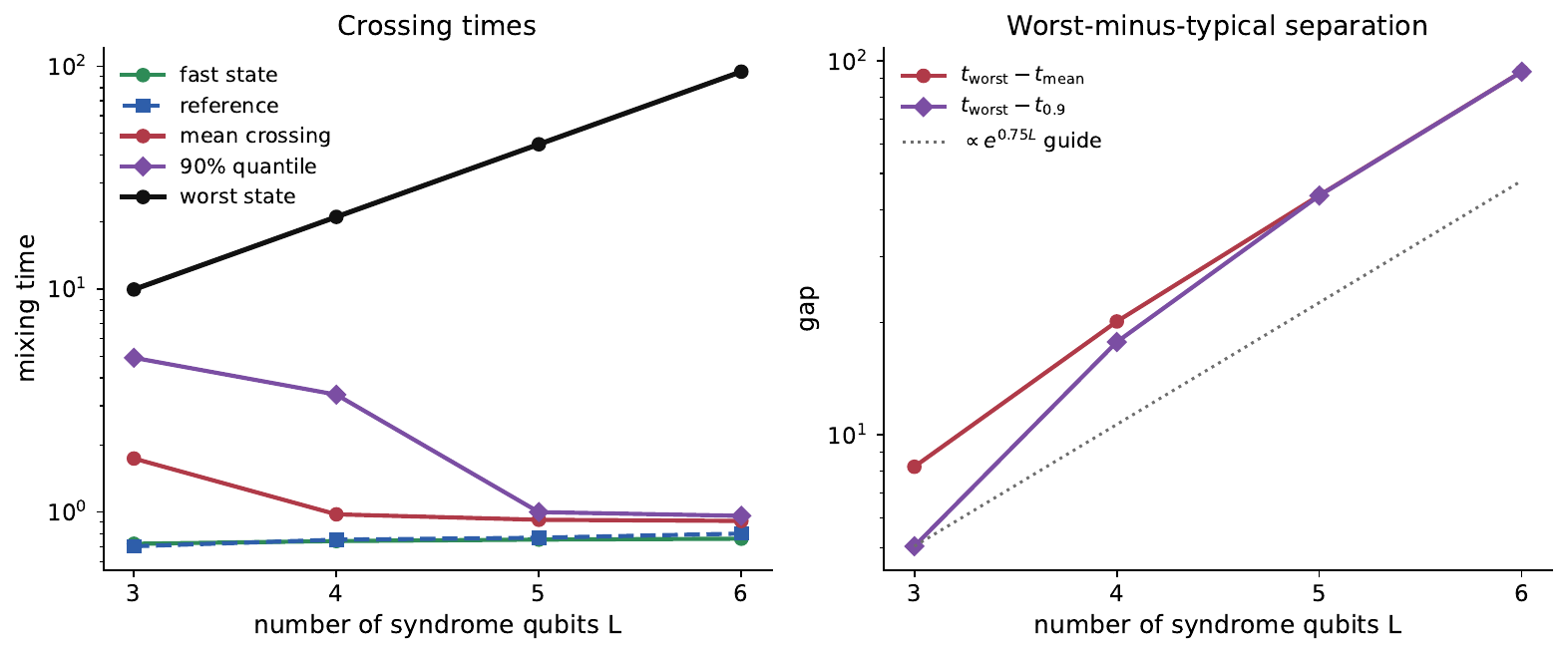}
	\caption{\textbf{System-size growth of the logical-sector separation.} Same microscopic model \eqref{eq:logicalmicro-L} for \(L=3,\dots,6\) at fixed threshold \(\eps=0.35\). The left panel compares the crossing times of the fast benchmark, the reference trajectory, the empirical mean, the empirical \(90\%\)-quantile, and the worst-case curve. The right panel shows the corresponding worst-minus-typical gaps. The reference and high-probability scales remain \(O(1)\), while the worst-case scale grows rapidly with \(L\) because it follows the exponentially weak logical relaxation rate \(\eta_L=e^{-cL}\).}
	\label{fig:logicalscaling}
\end{figure*}

\subsection{Microscopic Local-Qubit Realization and Numerical Settings}
\label{app:logicalsectornumerics}

For the figures we use a local qubit version of the product model. This part is only a numerical realization. It checks that the typical-fast and worst-slow split survives when the syndrome reset is written as local one-qubit thermalization.

We fix \(D_{\mathrm{log}}=2\). The logical register is one qubit, and the syndrome register consists of \(L\) physical qubits. For each \(L\),
\begin{equation}
	\mcH_L=\mathbb{C}^2_{\mathrm{log}}\otimes (\mathbb{C}^2)^{\otimes L}_{\mathrm{syn}},
	\qquad
	\sigma_L=
	\frac{\Id_{\mathrm{log}}}{2}\otimes \tau_\beta^{\otimes L},
	\qquad
	\tau_\beta=
	\mathrm{diag}(1-p_\beta,p_\beta),
	\qquad
	p_\beta=\frac{e^{-\beta}}{1+e^{-\beta}},
	\label{eq:logicalmicro-sigma}
\end{equation}
and
\begin{equation}
	\mcL_L^{\mathrm{micro}}
	=
	\eta_L\,\mathcal{D}_{\mathrm{log}}
	+
	\Gamma\sum_{j=1}^{L}\mathcal{R}^{(\beta)}_j,
	\qquad
	\eta_L=e^{-cL}.
	\label{eq:logicalmicro-L}
\end{equation}
Here \(\mathcal{D}_{\mathrm{log}}\) relaxes the logical qubit to \(\Id_{\mathrm{log}}/2\), while \(\mathcal{R}^{(\beta)}_j\) relaxes the \(j\)th syndrome qubit to \(\tau_\beta\) and acts trivially on all remaining tensor factors. In the figures we use
$
	\beta=1.8,
	\;
	\Gamma=2.0,
	\;
	c=0.75.
$
The two timescales are built directly into \eqref{eq:logicalmicro-L}. The syndrome register thermalizes on the \(O(\Gamma^{-1})\) scale, while the logical sector relaxes only through the exponentially small rate \(\eta_L=e^{-cL}\).

Because all reset generators commute, the finite-time channel factorizes. For the system sizes used below, every relaxation curve can therefore be evaluated exactly from the full finite-time map. No Trotter approximation or stochastic trajectory sampling is needed.

The reference trajectory in Fig.~\ref{fig:logicalbundle} is generated by \(\rho_{\mathrm{ref}}=\Id/d\). The slow benchmark is the logical-basis product state
\begin{equation}
	\rho_{\mathrm{slow}}
	=
	|0\rangle\langle 0|_{\mathrm{log}}
	\otimes
	|0\cdots 0\rangle\langle 0\cdots 0|_{\mathrm{syn}},
	\label{eq:logicalmicro-slow}
\end{equation}
and the fast benchmark is built from
\begin{equation}
	|\psi_{\mathrm{fast}}\rangle
	=
	\frac{
		|0,0,0,\dots,0\rangle
		+
		|1,1,0,\dots,0\rangle
	}{\sqrt{2}},
	\qquad
	\rho_{\mathrm{fast}}
	=
	|\psi_{\mathrm{fast}}\rangle\langle\psi_{\mathrm{fast}}|.
	\label{eq:logicalmicro-fast}
\end{equation}
The first entry refers to the logical qubit, and the remaining entries refer to the syndrome register. The logical reduced state of \(\rho_{\mathrm{fast}}\) is already maximally mixed. The slow logical contribution is therefore suppressed at \(t=0\), so the evolution is dominated by the fast syndrome relaxation.

To expose the same mechanism directly, we also monitor the logical-overlap diagnostic
\begin{equation}
	\Delta_{\mathrm{log}}(\psi)
	:=
	\left\|
	\rho_{\mathrm{log}}(\psi)-\frac{\Id_{\mathrm{log}}}{2}
	\right\|_1,
	\qquad
	\rho_{\mathrm{log}}(\psi)
	:=
	\Tr_{\mathrm{syn}}\!\bigl(|\psi\rangle\langle\psi|\bigr).
	\label{eq:logicalmicro-diagnostic}
\end{equation}
For the qubit specialization of \eqref{eq:logicalsector-alpha}, the corresponding comparison scale is
\begin{equation}
	\alpha_L(\delta)
	=
	\sqrt{\frac{3}{\delta(2^{L+1}+1)}}.
	\label{eq:logicalmicro-alpha}
\end{equation}
Figure~\ref{fig:logicaloverlap} shows that the empirical logical-overlap quantiles follow this suppression with system size.

For Fig.~\ref{fig:logicalbundle}, we use \(L=6\), \(d=128\), \(\eps=0.35\), \(64\) Haar-random pure states on \(\mcH_L\), and \(96\) log-spaced times. For Fig.~\ref{fig:logicalscaling}, we use \(L=3,4,5,6\), the same threshold \(\eps=0.35\), \(36\) Haar samples per size, and \(88\) log-spaced times per size. These settings resolve the two-scale structure. The reference and high-probability crossings stay \(O(1)\), while the slow benchmark follows \(\eta_L^{-1}\).

\section{Fixed-Rate Baselines and the Local Scope of Theorem~2}
\label{app:skinclosing}

The skin and boundary examples are kept at fixed rate. In these two branches the gap is meant to come from the overlap ratio, not from an additional closing of \(\gamma_2(L)\). If the rate also closes, the one-mode tail is multiplied by \(\gamma_2(L)^{-1}\), and the absolute gap no longer isolates the overlap geometry.

The connection with Theorem~2 is local in the threshold. For a given \(\eps\), the relevant quantity is the slope of \(\mu_L(t)\) at \(t_{*,L}(\eps)\). A small threshold may cross in the final one-mode tail. A moderate threshold may cross earlier, where subleading modes are still visible. The theorem only uses the slope in that crossing regime.

To make this precise, suppose that for a family indexed by \(L\) the mean curve enters a one-mode tail near the threshold \(\eps\),
\begin{equation}
	\mu_L(t)\approx A_L e^{-\gamma_2(L)t},
	\qquad
	A_L>0.
	\label{eq:skinclosing-tail}
\end{equation}
Then the deterministic crossing time satisfies
\begin{equation}
	t_{*,L}(\eps)\approx \frac{1}{\gamma_2(L)}\log\!\frac{A_L}{\eps},
	\label{eq:skinclosing-tstar}
\end{equation}
and the local crossing slope is
\begin{equation}
	m_L^{\times}(\eps)=-\mu_L'\bigl(t_{*,L}(\eps)\bigr)\approx \eps\gamma_2(L).
	\label{eq:skinclosing-slope}
\end{equation}
Let \(V_L^{\times}(\eps)\) denote the fixed-time vertical width of the bundle at the mean crossing and let \(H_L(\eps)\) denote the induced horizontal width of the threshold mixing-time distribution. Theorem~2 then gives the family-level estimate
\begin{equation}
	H_L(\eps)\approx \frac{V_L^{\times}(\eps)}{m_L^{\times}(\eps)}
	\approx
	\frac{V_L^{\times}(\eps)}{\eps\gamma_2(L)}.
	\label{eq:skinclosing-window}
\end{equation}
Equation~\eqref{eq:skinclosing-window} is only a crossing-regime estimate. It does not say that the ultimate tail controls every threshold. In thermodynamic families, the useful question is where the chosen threshold crosses and how steep the mean curve is there.

A closing rate weakens this local inverse step. When \(\gamma_2(L)=O(1)\), the horizontal width is set by the vertical concentration scale \(V_L^{\times}(\eps)\). When \(\gamma_2(L)\to0\), the same vertical band is magnified by \(\gamma_2(L)^{-1}\).

This is why Secs.~\ref{app:skin} and \ref{app:boundary} are kept at fixed rate. In the one-particle skin case, the logarithmic gap comes from the overlap geometry and the crossing slope remains \(O(\eps)\). In the boundary-supported many-body case, the linear branch comes from the exponential suppression of a typical boundary overlap. Neither baseline uses rate closing as the source of the gap.

If the skin rate is allowed to close with \(L\), the interpretation changes. For a polynomial closing \(\gamma_2(L)\sim L^{-\alpha}\), equation \eqref{eq:skinclosing-window} gives
\begin{equation}
	H_L(\eps)\approx \frac{L^{-1/2}}{\eps\gamma_2(L)}
	\sim
	\frac{1}{\eps}L^{\alpha-\frac{1}{2}}.
	\label{eq:skinclosing-poly}
\end{equation}
The theorem-level regime shrinks only when \(\alpha<1/2\). At \(\alpha=1/2\) the bound stops improving with \(L\). For \(\alpha>1/2\) it grows. For an exponential closing \(\gamma_2(L)\sim e^{-\kappa L}\), one obtains
\begin{equation}
	H_L(\eps)\approx \frac{e^{\kappa L}}{\eps\sqrt{L}},
	\label{eq:skinclosing-exp}
\end{equation}
so the local inverse step behind Theorem~2 is lost at the family level.

The same conclusion appears from the rare-state side. In the one-mode regime,
\begin{equation}
	\tworst(\eps)-t_{\mathrm{typ}}^{(\delta)}(\eps)
	\gtrsim
	\frac{1}{\gamma_2(L)}
	\log\!\frac{\|L_2\|_\infty}{\alpha_{\mathrm{typ}}(\delta)}.
	\label{eq:skinclosing-bottleneck}
\end{equation}
For the skin example the logarithmic factor grows as \(\log L\). If \(\gamma_2(L)\) also closes, the larger absolute separation is dominated by the common factor \(\gamma_2(L)^{-1}\). The gap then no longer isolates the extremal-to-typical overlap ratio that the fixed-rate skin example is meant to display.

There is no contradiction with Theorem~2. The theorem is a finite-size statement. What can fail is a uniform family-level use of the theorem when the crossing slope collapses with \(L\).

\begin{figure*}[t]
	\centering
	\includegraphics[width=0.90\textwidth]{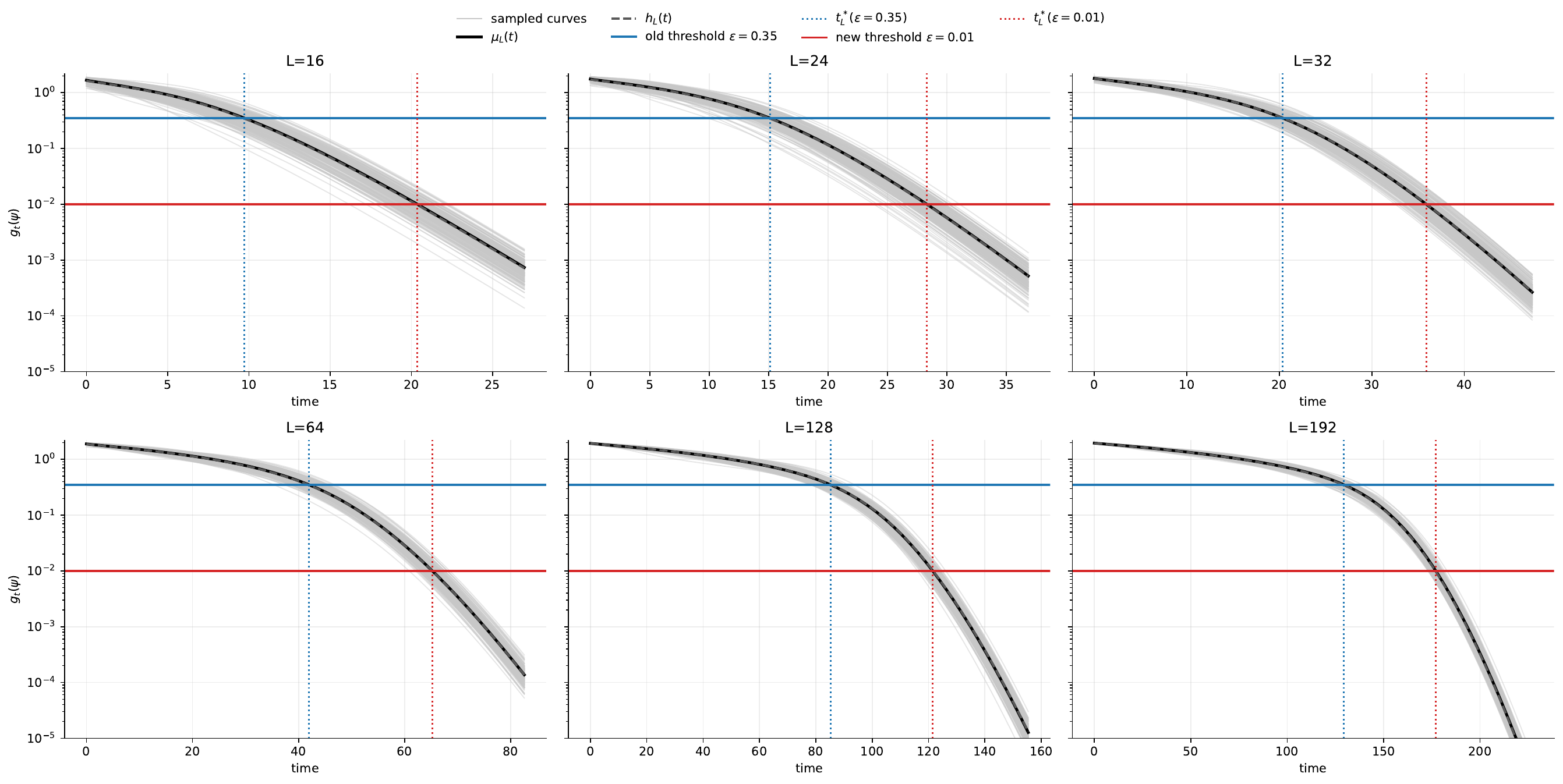}
	\caption{\textbf{Fixed-rate single-particle skin bundles in the strong regime.} Semilogy plots of the population-sector relaxation curves for the biased hopping chain with \((\gamma_R,\gamma_L)=(1.6,0.4)\), \(\lambda_L=1\), and \(\gamma_\phi=1\). The six panels correspond to \(L=16,24,32,64,128,192\). Gray curves are Haar-sampled trajectories \(g_t(\psi)\). The black curve is the empirical mean \(\mu_L(t)\). The blue dashed curve is the barycenter benchmark \(h_L(t)\). The blue horizontal line and blue dotted vertical line mark the threshold \(\eps=0.35\) and its deterministic crossing time \(t_L^*(0.35)\). The red horizontal line and red dotted vertical line mark the deeper threshold \(\eps=0.01\) and its deterministic crossing time \(t_L^*(0.01)\). Lowering the threshold moves the crossing deeper into the clean one-mode tail, while the higher-threshold regime still shows the same qualitative separation in the fixed-rate family.}
	\label{fig:skinclosing-bundle}
\end{figure*}

\begin{figure*}[t]
	\centering
	\includegraphics[width=0.96\textwidth]{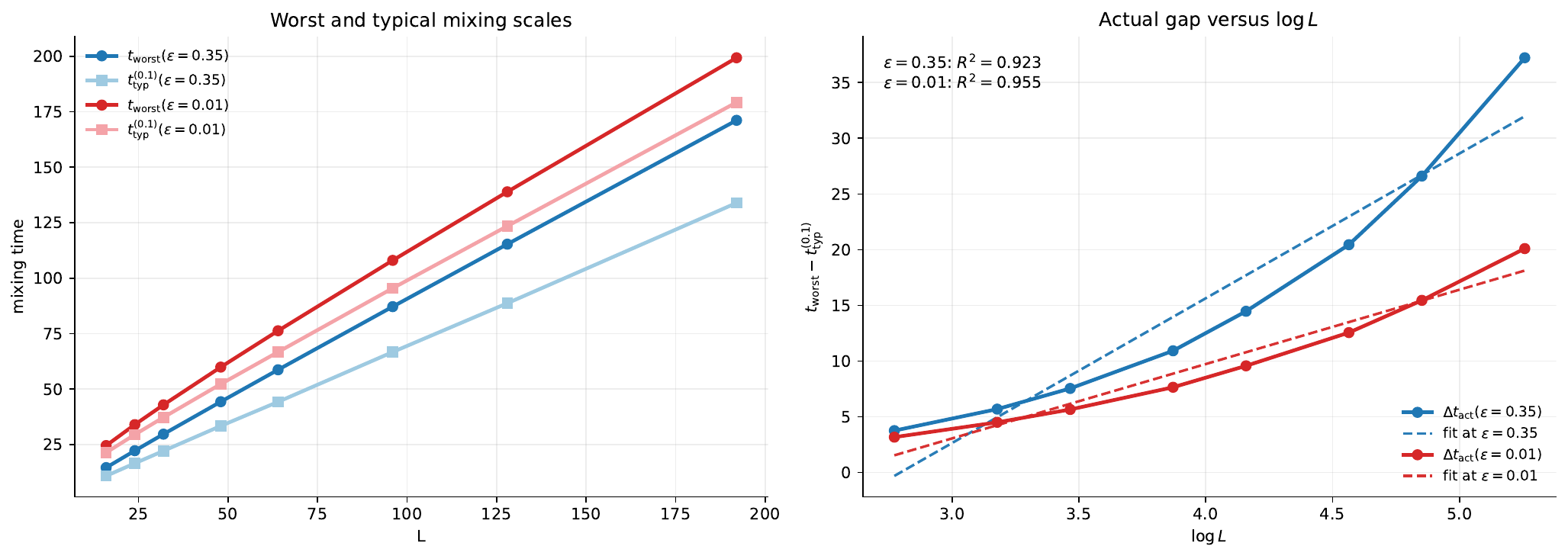}
	\caption{\textbf{Worst-versus-typical separation at two thresholds in the fixed-rate skin baseline.} Same strong-skin family as in Fig.~\ref{fig:skinclosing-bundle}. The left panel compares the worst-case mixing time with the empirical \(0.9\)-quantile \(q_{0.9}[\tmix(\psi,\eps)]\) for \(\eps=0.35\) and \(\eps=0.01\). The right panel plots the actual gap \(\Delta t_{\mathrm{act}}^{(0.1)}(L)=\tworst(\eps)-q_{0.9}[\tmix(\psi,\eps)]\) against \(\log L\), together with separate least-squares linear fits for the two thresholds. The lower threshold is closer to the one-mode tail and gives the cleaner logarithmic fit. The larger threshold still shows approximately logarithmic growth, with visible subleading-mode corrections near the crossing.}
	\label{fig:skinclosing-gap}
\end{figure*}

This separation of cases is consistent with the literature. Skin effects can slow relaxation without closing the gap \cite{Ueda_skin_effect}, while other boundary-dissipated localized models have exponentially closing gaps \cite{PhysRevB.106.064203}. The gap alone is not enough to identify the relevant relaxation law \cite{Mori20resolving,24prb_mori,PhysRevLett.134.140405}. We therefore keep the fixed-rate baselines separate from closing-rate stress tests.

The protected-sector branch is different. There the small rate \(\eta_L=e^{-cL}\) is part of the rare-sector mechanism itself. Typical states have negligible protected-sector weight and cross through the fast bulk. The worst-case trajectory follows the leakage time.

\subsection{Single-Particle Fixed-Rate Baseline}
\label{app:skinclosingnumerics}

Here we use the one-particle biased hopping chain in the population sector, so \(p(t)=e^{tQ_L}p(0)\) with open boundaries and right and left hopping rates \(\lambda_L\gamma_R\) and \(\lambda_L\gamma_L\). Throughout this subsection we fix the strong skin regime
\begin{equation}
	(\gamma_R,\gamma_L)=(1.6,0.4),
	\qquad
	\lambda_L=1,
	\qquad
	\gamma_\phi=1.
\end{equation}
We compare the original threshold \(\eps=0.35\) with the deeper threshold \(\eps=0.01\). Initial states are Haar-random pure states on the one-particle Hilbert space, reduced to populations by \(p_x(0)=|\psi_x|^2\). This reduction is sufficient for the present purpose because the added dephasing suppresses coherences faster than the population tail, so the late-time scale is governed by the same biased hopping sector. For the bundle plot we use \(200\) Haar samples for each \(L=16,24,32,64,128,192\). For the scaling plot we use \(500\) Haar samples for each \(L=16,24,32,48,64,96,128,192\). Within the classical population sector the worst case over the simplex is attained at a basis state, so the reported \(\tworst\) is obtained exactly by scanning the \(L\) basis populations.

Figure~\ref{fig:skinclosing-bundle} shows the semilogy bundles for six representative sizes. The blue markers use \(\eps=0.35\), and the red markers use the deeper threshold \(\eps=0.01\). The tails stay in a fixed-rate regime over the displayed range. In the population generator, the slow spectral gap decreases only from about \(0.43\) at \(L=16\) to about \(0.20\) at \(L=192\), so the displayed family does not show gap closing. Lowering the threshold moves the crossing deeper into the late-time tail. The higher threshold cuts the same family earlier, where subleading modes are still visible.

Figure~\ref{fig:skinclosing-gap} then isolates the rare-state separation in the same fixed-rate family. The left panel compares the empirical \(0.9\)-quantile with the worst-case scale at both thresholds. The right panel plots the actual gap
\[
\Delta t_{\mathrm{act}}^{(0.1)}(L)
:=
\tworst(\eps)-q_{0.9}\!\bigl[\tmix(\psi,\eps)\bigr]
\]
against \(\log L\). At the lower threshold \(\eps=0.01\), the fit is cleaner because the crossing lies deeper in the one-mode tail. At the larger threshold \(\eps=0.35\), the gap still grows close to \(\log L\), although subleading modes remain active near the crossing. We read this as a fixed-rate skin separation with finite-size corrections away from the clean one-mode tail.

\begin{figure*}[t]
	\centering
	\includegraphics[width=0.96\textwidth]{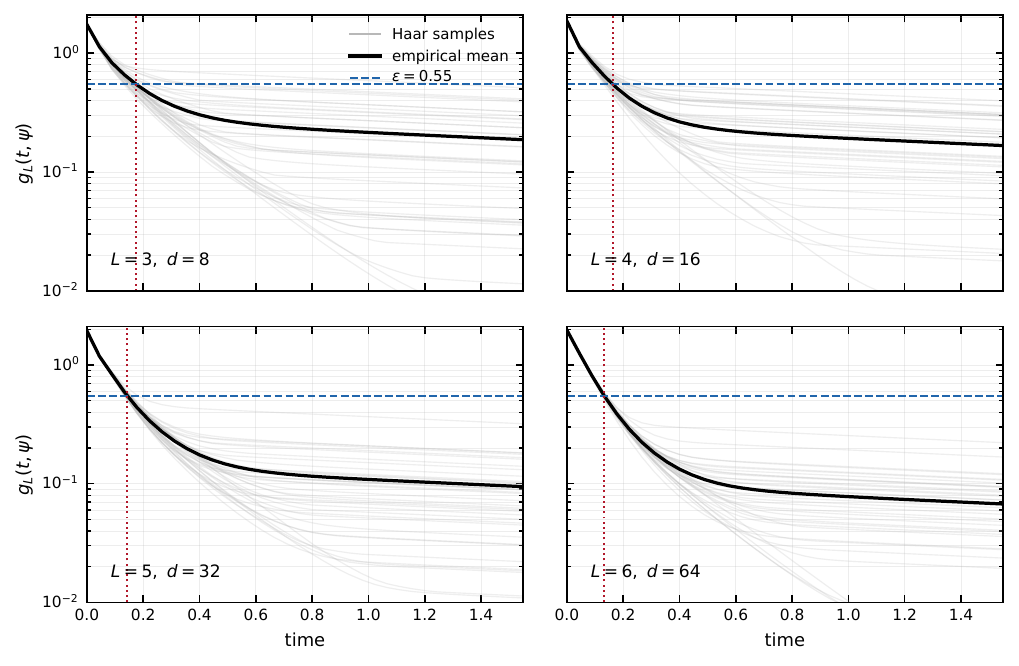}
	\caption{\textbf{Fixed-threshold bundles in the unital boundary-mode baseline.} Exact full-density trace-distance relaxation curves for the Pauli-Lindblad model \eqref{eq:boundarytoy-lindbladian} at \(\Delta=0.25\), \(\Gamma=4.0\), and fixed threshold \(\eps=0.55\). The panels show \(L=3,4,5,6\), with Hilbert-space dimensions \(d=8,16,32,64\). Gray curves are Haar-sampled trajectories \(g_L(t,\psi)\). The black curve is the empirical mean \(\mu_L(t)\). The blue dashed line marks the fixed threshold. The dotted vertical line marks the deterministic mean crossing. The curves become more tightly bundled with increasing \(d\), which gives a finite-size visualization of the fixed-threshold concentration mechanism.}
	\label{fig:boundary-toy-bundle}
\end{figure*}

\begin{figure*}[t]
	\centering
	
	\begin{minipage}[t]{0.485\textwidth}
		\centering
		\vspace{0pt}
		\includegraphics[width=\linewidth]{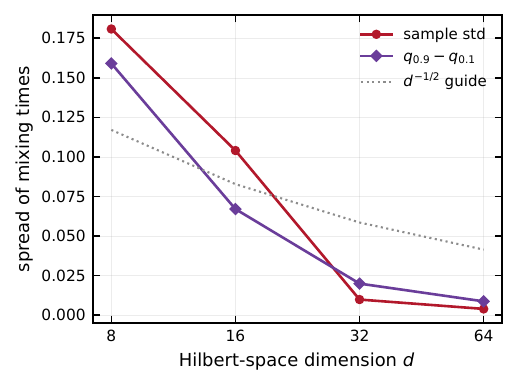}
	\end{minipage}
	\hfill
	\begin{minipage}[t]{0.485\textwidth}
		\centering
		\vspace{0pt}
		\includegraphics[width=\linewidth]{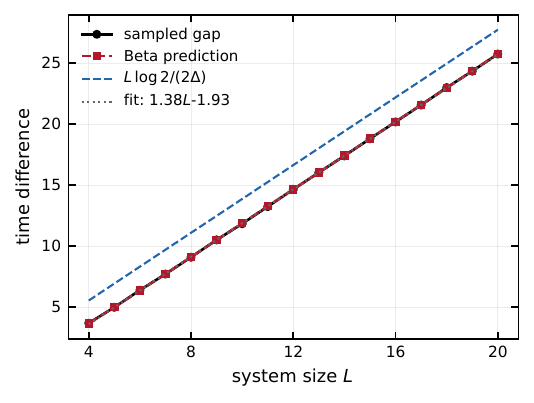}
	\end{minipage}
	
	\caption{\textbf{Fixed-threshold concentration and tail-window boundary separation.} Both panels use the unital boundary-mode baseline \eqref{eq:boundarytoy-lindbladian} with \(\Delta=0.25\) and \(\Gamma=4.0\). In the left panel, the threshold is fixed at \(\eps=0.55\). The sample standard deviation and the interquantile width \(q_{0.9}-q_{0.1}\) of \(\tmix(\psi,\eps)\) decrease with Hilbert-space dimension, showing the finite-size signature of horizontal concentration. In the right panel, the threshold is \(\eps_L=\eps_0 2^{-L/2}\) with \(\eps_0=0.10\). The black curve shows \(t_{\mathrm{worst}}^{\mathrm{basis}}(\eps_L)-q_{0.9}[\tmix(\psi,\eps_L)]\), the red curve shows the Beta-distribution prediction \eqref{eq:boundarytoy-prediction}, and the blue dashed line shows the asymptotic guide \(L\log 2/(2\Delta)\). The fixed-\(\eps\) panel tests concentration, while the scaled-\(\eps_L\) panel isolates the boundary-overlap bottleneck law.}
	\label{fig:boundary-toy-concentration-separation}
\end{figure*}

\subsection{Boundary-Supported Many-Body Fixed-Rate Baseline}
\label{app:skinclosingnumerics-manybody}

We now use a minimal unital Pauli-Lindblad baseline that exactly realizes the boundary-supported slow-mode hypothesis of Sec.~\ref{app:boundary}. The system is an \(L\)-qubit chain, \(d=2^L\), and
\begin{equation}
\begin{aligned}
	\mcL_L(\rho)
	=
	\frac{\Delta}{2}\bigl(X_1\rho X_1-\rho\bigr)
	+
	\frac{\Gamma}{2}\bigl(Z_1\rho Z_1-\rho\bigr) 
	+
	\frac{\Gamma}{2}\sum_{j=2}^{L}
	\bigl[
	X_j\rho X_j-\rho
	+
	Z_j\rho Z_j-\rho
	\bigr].
\end{aligned}
\label{eq:boundarytoy-lindbladian}
\end{equation}
Here \(X_j,Y_j,Z_j\) are Pauli operators on site \(j\). In the numerics we use
$
	\Delta=0.25,
	\;
	\Gamma=4.0.
$
All jump operators in \eqref{eq:boundarytoy-lindbladian} are Hermitian Pauli operators. The channel is therefore unital and has the unique stationary state
	$\sigma_L=\frac{\Id}{2^L}.$

The Heisenberg-picture action is diagonal in the Pauli-string basis. On the first site the nonzero single-site decay rates are
\begin{equation}
	r_1(Z)=\Delta,
	\qquad
	r_1(X)=\Gamma,
	\qquad
	r_1(Y)=\Gamma+\Delta,
\end{equation}
while on each bulk site \(j\ge 2\),
	$r_j(X)=r_j(Z)=\Gamma,
	\;
	r_j(Y)=2\Gamma.$
For \(0<\Delta<\Gamma\), the unique slowest nonstationary left eigenoperator is
\begin{equation}
	L_2=Z_1\otimes \Id_{2\cdots L},
	\qquad
	\Tr(L_2)=0.
	\label{eq:boundarytoy-slowmode}
\end{equation}
With \(R_2=L_2/d\), one has \(\|R_2\|_1=1\) and \(\|L_2\|_\infty=1\). Thus \eqref{eq:boundarytoy-lindbladian} is an exact fixed-rate realization of the boundary-supported setting in Sec.~\ref{app:boundary}, with \(q=2\), \(\ell=1\), and \(\gamma_2=\Delta\).

The two threshold choices below test different parts of the theory. At a fixed threshold, we inspect the concentration of full trace-distance mixing times. This is the same vertical-to-horizontal mechanism as Theorem~2. It is not meant to test the linear one-mode bottleneck law. At fixed \(\eps\), the Haar-typical slow overlap eventually becomes smaller than the threshold itself, so the typical crossing need not occur in the final one-mode tail.

To test the boundary-overlap bottleneck law, we use the scaled threshold
\begin{equation}
	\eps_L=\eps_0 2^{-L/2},
	\qquad
	\eps_0=0.10.
	\label{eq:boundarytoy-scaled-threshold}
\end{equation}
This choice keeps the typical slow-overlap crossing inside the one-mode tail along the finite-size sequence. The threshold dependence then cancels from the time difference, and the leading growth comes only from the ratio between the worst and typical overlaps.

For a Haar-random pure state, the slow overlap is
	$a_2(\psi)=\langle \psi|Z_1|\psi\rangle.$
If
	$B_L(\psi)=\sum_{x:x_1=0}|\psi_x|^2,$
then
	$B_L(\psi)\sim \mathrm{Beta}(2^{L-1},2^{L-1}),\;
	a_2(\psi)=2B_L(\psi)-1.$
The computational-basis worst state has \(|a_2|=1\). In the one-mode tail,
	$g_L(t,\psi)\simeq |a_2(\psi)|e^{-\Delta t}.$
The overlap-quantile prediction for the worst-minus-typical gap is therefore
\begin{equation}
	\Delta t_{\mathrm{pred}}^{(0.1)}(L)
	=
	\frac{1}{\Delta}
	\log
	\frac{1}{q_{0.9}(|a_2|)}.
	\label{eq:boundarytoy-prediction}
\end{equation}
Since \(q_{0.9}(|a_2|)=2^{-L/2}O(1)\), this gives
\begin{equation}
	\Delta t_{\mathrm{pred}}^{(0.1)}(L)
	=
	\frac{L\log 2}{2\Delta}
	+
	O(1).
	\label{eq:boundarytoy-linear}
\end{equation}

Figure~\ref{fig:boundary-toy-bundle} shows the full-density relaxation bundles at fixed threshold. The data use Haar-random pure states and the exact finite-time Pauli channel, with no Trotter approximation or trajectory sampling. The fixed-threshold choice \(\eps=0.55\) keeps the crossing in the regular part of the relaxation curve for these sizes. The figure is therefore a concentration check rather than a one-mode bottleneck check.

The left panel of Fig.~\ref{fig:boundary-toy-concentration-separation} turns the same fixed-threshold data into mixing-time statistics. In the displayed run, the sample standard deviation decreases from about \(0.181\) at \(d=8\) to about \(0.0039\) at \(d=64\), and the interquantile width \(q_{0.9}-q_{0.1}\) decreases from about \(0.159\) to about \(0.0087\). The mean crossing remains \(O(1)\). This is the finite-size horizontal concentration expected from Theorem~2.

The right panel of Fig.~\ref{fig:boundary-toy-concentration-separation} tests the one-mode boundary-overlap law. The scaled threshold \eqref{eq:boundarytoy-scaled-threshold} makes the typical slow overlap and the threshold comparable, so both the worst and \(0.9\)-quantile crossings occur in the same slow tail. The sampled gap agrees with the Beta prediction \eqref{eq:boundarytoy-prediction}. A least-squares fit gives a slope about \(1.38\), close to the theoretical value
	$\frac{\log 2}{2\Delta}=1.386.$
Thus the growing separation in this baseline is not caused by a closing gap. It comes from the suppression \(q_{0.9}(|a_2|)=2^{-L/2}O(1)\) of the Haar-typical boundary overlap, exactly as in Sec.~\ref{app:boundary}.

\bibliography{refs}